%% file: main.tex
\documentclass[a4paper,11pt]{article}
\pdfoutput=1 
\usepackage{jcappub} 
\usepackage[T1]{fontenc}
\usepackage[nameinlink,noabbrev]{cleveref}
\usepackage[dvipsnames]{xcolor}
\usepackage[normalem]{ulem}
\usepackage{mathrsfs}
\usepackage{comment}
\usepackage{makecell, multirow}
\usepackage{orcidlink}
\usepackage{booktabs}
\usepackage{subcaption}
\usepackage{lineno}
\usepackage{acronym}
\usepackage{url}
\input{acronyms}

\definecolor{brick}{HTML}{88422B}

\title{\textbf{Cosmology beyond standard sirens: cross-correlation of gravitational waves and neutral hydrogen intensity mapping}}

\author[1,2]{Matteo Schulz\orcidlink{0009-0005-8184-0232},}
\author[3]{Ulyana Dupletsa\orcidlink{0000-0003-2766-247X},}
\author[1,2]{Andrea Cozzumbo\orcidlink{0009-0004-2772-2692},}
\author[4]{Giona Sala\orcidlink{0009-0001-3716-862X},}
\author[5,6]{Tommaso Ronconi\orcidlink{0000-0002-3515-6801},}
\author[7,8]{Marta Spinelli\orcidlink{0000-0003-0148-3254},}
\author[9,10,11]{Riccardo Murgia\orcidlink{0000-0002-2224-7704},}
\author[12]{Simone Mastrogiovanni\orcidlink{0000-0003-1606-4183},}
\author[1,2]{Jan Harms\orcidlink{0000-0002-7332-9806},}
\author[4]{Erik de la Haye\orcidlink{0009-0000-9433-1684},}
\author[13]{Felicitas Keil\orcidlink{0000-0002-8108-1679
}}

\affiliation[1]{Gran Sasso Science Institute (GSSI), Viale F. Crispi 7, L'Aquila (AQ), I-67100, Italy}
\affiliation[2]{INFN - Laboratori Nazionali del Gran Sasso (LNGS), L'Aquila (AQ), I-67100, Italy}
\affiliation[3]{Marietta Blau Institute - Austrian Academy of Sciences, 1010 Vienna, Austria}
\affiliation[4]{
Institute for Theoretical Particle Physics and Cosmology, RWTH Aachen University,
D-52056 Aachen, Germany}
\affiliation[5]{INAF - Institute of Radioastronomy (IRA), Via Gobetti 101, 40129 Bologna, Italy}
\affiliation[6]{INFN, Sezione di Bologna, Viale Carlo Berti Pichat, 6/2, 40127 Bologna BO}
\affiliation[7]{Observatoire de la Côte d’Azur, Laboratoire Lagrange, Bd de l’Observatoire, 06304 Nice, France}
\affiliation[8]{Department of Physics \& Astronomy, University of the Western Cape, Cape Town 7535, South Africa}
\affiliation[9]{Dipartimento di Fisica, Universit\`a degli Studi di Cagliari, Cittadella Universitaria, 09042 Monserrato (CA), Italy}
\affiliation[10]{INFN, Sezione di Cagliari, Cittadella Universitaria, 09042 Monserrato (CA), Italy}
\affiliation[11]{INAF - Osservatorio Astronomico di Cagliari, Via della Scienza 5, 09047 Selargius (CA), Italy}
\affiliation[12]{INFN, Sezione di Roma, 1-00185 Roma, Italy}
\affiliation[13]{Université de Toulouse, CNRS, IRAP, CNES, 14 Av. Édouard Belin, 31400 Toulouse, France}

\emailAdd{matteo.schulz@gssi.it}

\abstract{
We explore the potential of cross-correlation between gravitational wave (GW) events and neutral hydrogen (\textrm{H\textsc{i}}) intensity mapping surveys to serve as an independent cosmological probe.
Focusing on the \ac{et} and the \ac{skao}, and assuming that binary black hole mergers and \textrm{H\textsc{i}} emission are biased tracers of the underlying dark matter distribution, we use their angular auto- and cross-correlation spectra to constrain cosmological parameters. 
We test three different GW detector networks: \ac{et} alone, both in its $\Delta$ and 2L configuration, and ET-2L together with Cosmic Explorer.
We show that the cross-correlation method, by naturally mitigating tracer-specific systematics, yields robust cosmological bounds, allowing for a sub-percent ($\sim 0.5\%$) precision on the Hubble constant $H_\mathrm{0}$.
Furthermore, this approach robustly constrains the cosmic expansion history throughout the post-reionization era of the Universe and, unlike conventional standard sirens, simultaneously probes the large scale distribution of matter perturbations, achieving relative uncertainties of approximately 1.3\% on the total matter density $\Omega_\mathrm{m}$ and 1.6\% on the late-time clustering amplitude $\sigma_8$.
}

\begin{document}
\maketitle
\flushbottom
\newpage

\acresetall

\section{Introduction}
\label{sec:intro}
Driven by the discovery of the accelerated expansion of the Universe~\cite{Riess1998Observational, Perlmutter1998Measurements}, several cosmological probes have been thoroughly studied and improved in the last twenty years, such as the \ac{cmb}~\cite{Aghanim2018PlanckVI}, the Supernovae Type Ia~\cite{Riess2019Large}, and the Baryon Acoustic Oscillations (BAOs)~\cite{Font_Ribera_2014}, becoming standard probes in the landscape of modern cosmology. 
As they reached percent/sub-percent accuracy, disagreements between early- and late-universe measurements of some cosmological parameters~\cite{Verde2019Tensions} have emerged, such as the so-called Hubble constant ($H_\mathrm{0}$) tension~\cite{DiValentino2021Realm, kamionkowski2022hubbletensionearlydark, hu2023hubbletensionevidencenew, Poulin2018Early}.
In light of this, complementary and independent probes are needed to test the Universe~\cite{Moresco:2022phi, Abbott2017GW170817, Wong2019H0LiCOW, Freedman2019Carnegie}.

Traditionally, measuring cosmological distances relies on the cosmic distance ladder, a series of sequentially calibrated methods built upon ``standard candles'', which are astrophysical objects of known intrinsic luminosity~\cite{Riess2019Large}.
In contrast, \ac{gw} events can serve as ``standard sirens''.
Since they provide a direct measurement of the luminosity distance of the source from first principles of General Relativity~\cite{schutz_1986}, \acp{gw} require no other form of calibration, making them independent of the cosmic distance ladder.
In order to constrain the cosmic expansion history using \acp{gw} as standard sirens, additional information on the redshift of the source is needed.
However, redshift cannot be inferred from \ac{gw} data alone, due to the existing degeneracy between redshift and the measured source mass.
Breaking this degeneracy is required to obtain information on both cosmological parameters and \ac{bbh} population parameters (e.g., their mass distribution)~\cite{Ezquiaga:2022zkx}.

Several different techniques have been investigated to break the degeneracy between \ac{gw} parameters~\cite{Chen:2024gdn}. 
These include the use of electromagnetic counterparts from ``bright sirens''~\cite{Markovic_1993cr, dalal_2006, LIGOScientific:2017adf, Feeney:2018mkj, Palmese:2020, Mancarella:2024qle, cozzumbo2024}, matching the \ac{gw} localization with galaxy catalogs in the ``dark sirens'' method~\cite{schutz_1986, Holz:2005df, MacLeod:2007jd, DelPozzo:2011vcw, Nishizawa:2016ood, Chen:2017rfc, LIGOScientific:2018gmd, DES:2019ccw, Gray:2019ksv, LIGOScientific:2019zcs, Palmese:2020, Finke:2021aom, Abbott_2023, Mancarella_2022_physRevD, Gair:2022zsa, Gray:2023wgj, Borghi:2023opd, Bom:2024afj, Borghi:2025nis, lvk2026gwtc50_cosmo}, and exploiting features in the \ac{gw} mass distribution for ``spectral sirens'' \cite{Chernoff:1993th, Taylor:2012db, You:2020wju, Ye:2021klk, Ezquiaga:2022zkx, Mastrogiovanni:2023emh, Ferraiuolo:2025evh, MaganaHernandez:2025cnu, Pierra:2026ffj, Tagliazucchi:2026dpr, Bertheas:2026odj,lvk2026gwtc50_cosmo}.
In this work, we adopt the ``cross-correlation method''~\cite{Diaz:2021pem, Scelfo:2021fqe, Mukherjee:2022afz, Mali:2024wpq, Ferri:2024amc, pedrotti2025, sala2025, dematos2025, Dalang:2024gfk, Pan:2025iya, Cross-Parkin:2026wyz}, which is based on the assumption that both \ac{gw} sources and the galaxy population are tracers of the \ac{lss}~\cite{Oguri_2016, Bera:2020, mukherjee2018}.
This approach relies on the spatial clustering between \ac{gw} events and \ac{lss} tracers.
Specifically, their angular cross-correlation signal peaks when the assumed cosmological model correctly maps the \ac{gw} luminosity distances to the redshifts of the \ac{lss} catalog.
While constraining the luminosity distance-redshift relation is crucial for probing the background expansion history of the Universe, yielding bounds on the Hubble constant $H_\mathrm{0}$ and the matter abundance $\Omega_\mathrm{m}$, the cross-correlation signal is also sensitive to the linear evolution of matter density perturbations.
This allows to determine the amplitude of the matter power spectrum, making it possible to probe both the expansion of the Universe and the growth of structures.
A major advantage of the cross-correlation method is its intrinsic robustness to observational and theoretical systematics.
Because it relies on the statistical combination of independent datasets, it naturally cancels out tracer-specific errors, making the resulting cosmological inference highly resilient against unmodeled systematics.
Compared to standard siren techniques, which, within the standard cosmological model, are only sensitive to the background quantities, i.e. the current expansion rate of the Universe $H_\mathrm{0}$ and the matter density $\Omega_\mathrm{m}$, the cross-correlation signal simultaneously probes the growth of matter perturbations.
This makes it sensitive to the linear amplitude of density fluctuations, $\sigma_8$.

In this context, line \ac{im}~\cite{ kovetz2017lineintensitymapping2017status, Bernal_2022, kovetz2019astrophysicscosmologylineintensitymapping} has emerged as a promising probe to trace the \ac{lss} distribution and provide stringent cosmological constraints~\cite{santos2015cosmologyskahiintensity, Bernal_2019}.
This technique allows for a very precise redshift measurement of the sources, since the observed wavelength of a known emission line with respect to the observed wavelength directly maps their redshift along the line of sight.
Among the different targets for \ac{im}, the redshifted 21~cm line of \ac{hi} represents one of the most promising ones, and several detections of the signal in cross-correlation with galaxy surveys have already been achieved~\cite{chang_tc:2010, Masui_2013, anderson_2018, Wolz_2021, chimecollaboration_2023}.

In particular, a \ac{hi} \ac{im} survey carried out with the MeerKAT telescope~\cite{Wang_2021} has successfully achieved large-scale detections of the cross-correlation signal around $z \sim 0.4$, e.g.~\cite{meerKLASS_2025}.
Notably, a wide-area MeerKAT survey extending to $z \sim 1.5$ is ongoing~\cite{santos2017meerklassmeerkatlargearea}.
In the near future, the advanced capabilities of the forthcoming international project \ac{skao}~\cite{braun_2015} will be crucial for achieving wider and deeper \ac{im} surveys of the \ac{lss}, ultimately covering the entire post reionization Universe~\cite{santos2015cosmologyskahiintensity, Bacon_Battye_2020}.

The synergy between \ac{gw} events and \ac{hi} \ac{im} relies on the fact that these two probes are assumed to be \textit{biased} tracers of the same underlying \ac{dm} distribution.
Neutral hydrogen is the most abundant element in the Universe, and, as all baryonic matter, it accumulates within the gravitational potential well of \ac{dm} halos, fueling the star formation that then yields the stellar-mass \ac{bh} binaries, which produce \ac{gw} events.
Therefore, on cosmological scales, the spatial clustering of \ac{gw} sources is expected to follow the same distribution of \ac{hi} emission, thus resulting in a non-zero cross-correlation signal.

In this work, we investigate the synergy between next-generation observatories by characterizing the auto- and cross-correlation signal between \ac{gw}s and \ac{hi}, focusing on \ac{skao} for \ac{hi} \ac{im}, and \ac{et}~\cite{puntuto_2010, Punturo_2022, Maggiore_2020, Abac_2026} for \ac{gw}s.
To assess the impact on the constraining power due to different \ac{gw} detector configurations, we perform the analysis across three different \ac{gw} networks: \textit{(i)} ET in its $\Delta$ configuration (ET-$\Delta$); \textit{(ii)} \ac{et} in its 2L configuration (ET-2L); and \textit{(iii)} ET-2L together with one Cosmic Explorer (CE) of 40 km (ET2L+CE).
For further details on the specific properties of the different setups refer to \cite{Branchesi_2023}.

The paper is organized as follows: Section~\ref{sec:methods} details the theoretical framework for the angular cross-correlation spectrum and the methodology used, Section~\ref{sec:tracers} describes the \ac{gw} and \ac{hi} tracers according to \ac{et} and \ac{skao} capabilities, Section~\ref{sec:results} presents the results of our forecasts and the impact of the different runs, and Section~\ref{sec:conclusions} summarizes our conclusions and outlines future prospects.

\section{Synergy of Gravitational Waves and Neutral Hydrogen}\label{sec:methods}
The cross-correlation between \ac{gw}s and \ac{hi} \ac{im} offers a unique opportunity for precision cosmology.
The physical relevance of this technique is based on the fact that both observables independently trace the same underlying \ac{lss} of the Universe.
Combining data from next-generation \ac{gw} observatories with \ac{skao} intensity maps allows one to infer the \ac{gw} redshift distribution and to constrain cosmology, as has been proven,~e.g., in \cite{Scelfo:2021fqe} and references therein.
Building on this synergy, the recent ``Radio Sirens'' methodology~\cite{dupletsa2026} highlighted how 3-dimensional \ac{hi} density fields can serve as high-resolution redshift priors for \ac{gw} events, yielding high-precision measurements of the Hubble constant, $H_\mathrm{0}$, fully complementary with galaxy surveys information.
While the approach developed in \cite{dupletsa2026} considers the full information from every single \ac{gw} source, the cross-correlation method relies on the use of summary statistics from the resolved \ac{bbh} sources.
This enables us to greatly reduce the computing cost of the analysis and consider even the weaker \ac{gw} signals that would be seen by next-generation \ac{gw} detectors.

One key strength of the cross-correlation method is its intrinsic robustness to both observational and theoretical systematics.
Since this technique relies on the statistical combination of physically independent datasets, such as \ac{gw} events and \ac{hi} \ac{im}, tracer-specific errors do not cross-correlate.
As a result, uncorrelated instrumental noise, unmodeled theoretical systematics, and foreground residuals naturally vanish from the cross-spectrum, thus yielding a robust cosmological inference~\cite{switzer_2013}.
Furthermore, unlike conventional standard siren techniques, which are limited to constraining only background cosmological parameters (i.e. the cosmic expansion rate), the auto- and cross-correlation spectra trace the actual growth and distribution of cosmic structures.
Therefore, given the additional sensitivity of this method to matter fluctuations, this allows us to infer not only the background expansion history, but also the amplitude of matter fluctuations, enabling simultaneous constraints on the primordial power spectrum alongside the standard background expansion history~\cite{Scelfo:2021fqe}.

We will now describe in detail the statistical methods that we have adopted.
In Subsection~\ref{ang_spectrum} we describe the formalism of the angular power spectrum, in Subsection~\ref{sec:tracers} we discuss the specific case of the two tracers considered in this work, in Subsection~\ref{sec:likelihood} we present our likelihood for cosmological parameter estimation.

\subsection{Tomographic angular power spectra}\label{ang_spectrum}
The formalism to compute the cross-correlation signal between any pair of \ac{lss} tracers is well established~\cite{regos_1989, Scharf_1992, Lahav_1994, Fisher_1994}.
In the context of this work, it proves to be particularly powerful for multiple reasons: $(i)$ given the lack of electromagnetic counterparts for \ac{bbh} mergers, it allows for the calibration of the redshift distribution of \ac{gw} events thanks to the tomographic nature of the \ac{hi} \ac{im} signal~\cite{menard2014clustering}; $(ii)$ the cross-correlation of two independent probes provides intrinsic robustness to instrumental and observational systematics, as uncorrelated errors between the datasets of the different tracers naturally vanish in the cross-spectrum~\cite{switzer_2013}. 

For a given tracer \textit{X}, its matter density contrast $\delta_X$ at a generic position in the sky $(\theta, \phi)$ and radial distance $x$ can be described as
\begin{equation}
    \delta_X(\theta, \phi, x) = \frac{\rho_X(\theta, \phi, x) - \langle \rho_X \rangle(x)}{\langle \rho_X \rangle(x)} \, ,
\end{equation}
where $\rho_X (x,\theta,\phi)$ is the matter density and $\langle \rho_X \rangle(x)$ is the average matter density on the entire sky at radial distance $x$.
By projecting it onto the sky surface, it is possible to expand the matter fluctuations in spherical harmonics as
\begin{equation}
    \delta_X(\theta, \phi, x) = \sum_{\ell =0}^{+\infty} \sum_{m =-\ell}^{+\ell} a_{\ell m}^X(x)Y_{\ell m}(\theta, \phi) \, ,
\end{equation}
where $a_{\ell m}^X(x)$ and $Y_{\ell m}(\theta,\phi)$ represent, at each angular scale $\ell$ and mode $m$, the harmonic coefficients and the spherical harmonics, respectively.
The angular power spectrum $\tilde{C}_\ell^{XY} (x_i, x_j)$ between tracers $X$ and $Y$, at radial distances $x_i$ and $x_j$, is defined as the covariance of their angular harmonics coefficients
\begin{equation}
   \langle a_{\ell m}^X(x_i), a_{\ell' m'}^{Y^*}(x_j) \rangle = \delta_{\ell \ell'} \delta_{m m'} \tilde{C}_\ell^{XY} (x_i, x_j) \, ,
\end{equation}
with $\delta$ denoting Kronecker deltas.
Crucially, since the harmonic coefficients encode the decomposition onto spherical harmonics of the matter density contrast, the angular power spectrum $\tilde{C}_\ell$ directly measures the covariance of these projected fluctuations on the sky.

To explicitly describe the angular power spectrum, we can decompose the harmonic coefficients into two separate terms, one including the signal contribution $s_{\ell m}$ and one including the noise contribution $n_{\ell m}$
\begin{equation}
    a_{\ell m}^X(x_i) = s_{\ell m}^X(x_i) + n_{\ell m}^X(x_i) \, .
\end{equation}
Assuming that the signal and noise are statistically independent, their cross-terms are null, i.e. $\langle s_{\ell m}^X(x_i), n_{\ell' m'}^{Y^*}(x_j) \rangle = 0 $.
We can therefore express the non-vanishing terms as:
\begin{equation}
    \langle s_{\ell m}^X(x_i), s_{\ell' m'}^{Y^*}(x_j) \rangle = \delta_{\ell \ell'} \delta_{m m'} C_\ell^{XY} (x_i, x_j) \, ,
\end{equation}
\begin{equation}
    \langle n_{\ell m}^X(x_i), n_{\ell' m'}^{Y^*}(x_j) \rangle = \delta_{\ell \ell'} \delta_{m m'} \mathcal{N}_{\ell}^{XY} (x_i, x_j)\, ,
\end{equation}
where $C_\ell^{XY} (x_i, x_j)$ is the theoretical angular power spectrum and $\mathcal{N}_{\ell m}^{XY} (x_i, x_j)$ is the noise spectrum and represents the covariance of the projected noise fluctuations on the sky.
It is then possible to write the simulated\footnote{In this work, we will refer with $\sim$ to the spectrum obtained from simulated mock observations.} angular power spectrum as
\begin{equation}
    \tilde{C}_\ell^{XY} (x_i, x_j) = C_\ell^{XY} (x_i, x_j) + \mathcal{N}_{\ell}^{XY} (x_i, x_j) \, .
\end{equation}
The theoretical power spectrum describing the correlation between two tracers $(X,Y)$ at any distance $(x_i, x_j)$, is defined as
\begin{equation}
    C_\ell^{XY} (x_i, x_j) = \frac{2}{\pi} \int \frac{dk}{k} \mathcal{T}_{\ell,X}(k, x_i) \mathcal{T}_{\ell,Y}(k, x_j) \mathcal{P} \left( k \right) \, ,
    \label{eq:cl_th}
\end{equation}
where $k$ is the comoving wavenumber, $\mathcal{P}(k) = k^3 P(k)$ is the primordial matter power spectrum, and $\mathcal{T}_{\ell,X}(k, x_i)$ and $\mathcal{T}_{\ell,Y}(k, x_j)$ are the effective transfer functions for each tracer at a given distance, and are generically defined as
\begin{equation}
    \mathcal{T}_{\ell,X}(k, x_i) = \int_0^{\infty} dx \mathcal{W}(x, x_i, \Delta x_i) \Delta_{\ell,X}(k, x) \, .
    \label{eq:delta_transfer}
\end{equation}
Here, $\mathcal{W}(x, x_i, \Delta x_i)$ is the normalized window function at distance $x_i$ with half-width $\Delta x_i$:
\begin{equation}
    \mathcal{W}(x, x_i, \Delta x_i) = \frac{W(x, x_i, \Delta x_i) \frac{dN_X}{dx}}{\int dx' W(x', x_i, \Delta x_i) \frac{dN_X}{dx'}} \, ,
    \label{eq:unnorm_windowfunc}
\end{equation}
where $W(x, x_i, \Delta x_i)$ is the un-weighted window function and $\frac{dN_X}{dx}$ is the source number density per distance interval.
Note that the normalized window functions depend on the specific observational set-ups for the two tracers and include contributions from the observed redshift distribution of each tracer, redshift and distance measurement errors.
For this reason, an exact description of the distance distribution of each tracer is needed (see Subsection~\ref{sec:tracers}).
Moreover, $\Delta_{\ell,X}(k, x)$ is the angular number count fluctuation of the tracer X, which depends on a dominant density term $\Delta_{\ell,X}^\mathrm{den}(k, x)$, and subdominant\footnote{All other contributions beyond the primary density term have been shown to be subdominant~\cite{pedrotti2025}.} effects~\cite{Challinor_2011, Bonvin_2011}, namely: velocity\footnote{The velocity term includes both Doppler effects and Redshift/Luminosity distance Space Distortion.} $\Delta_{\ell,X}^\mathrm{vel}(k, x)$, lensing $\Delta_{\ell,X}^\mathrm{len}(k, x)$, and gravity $\Delta_{\ell,X}^\mathrm{gr}(k, x)$.
The full expression reads
\begin{equation}
    \Delta_{\ell,X}(k, x) = \Delta_{\ell,X}^\mathrm{den}(k, x) + \Delta_{\ell,X}^\mathrm{vel}(k, x) + \Delta_{\ell,X}^\mathrm{len}(k, x) + \Delta_{\ell,X}^\mathrm{gr}(k, x) \, .
    \label{eq:angularFluctuations}
\end{equation}
For the complete description of each term in Equation~\eqref{eq:angularFluctuations} please refer to Appendix~\ref{appendix:rel_n_counts}.

In order to trace the distance evolution of the cross-correlation signal, we apply the definition of Equation~\eqref{eq:cl_th} with a tomographic approach, by dividing the source distribution into different distance bins.
Such a tomographic slicing allows for the extraction of maximum constraining power from the simulated data, as it preserves the radial clustering information that would otherwise be washed out by projection effects.
Note that previous works rely on the Limber approximation~\cite{Limber1953} to model the angular power spectrum.
In this work, we instead take a step forward, we go beyond the Limber approximation, and  additionally adopt a fully relativistic treatment~\cite{Dio_2013} to compute the exact form of Equation~\eqref{eq:cl_th}.
To achieve this, we utilize a custom fork of \texttt{Multi\_CLASS}~\cite{Bellomo20_multiclass}, which implements the exact integration formalism from \cite{schoeneberg_cls}.
The code used for this analysis is publicly available on GitHub\footnote{\url{https://github.com/MatteoSchulz/GWxHI_MultiCLASS}}.

Given that the noise is intrinsically linked to the observational setup for each type of source, the noise spectrum has to be discussed separately for each tracer.
In the specific case of discrete sources, such as \ac{gw} signals, the dominant noise source is shot noise.
The general form of the noise spectrum for a resolved tracer $X$ is given by:
\begin{equation}
    \mathcal{N}_\ell^{XX} (x_i, x_j) = 4 \pi f^X_{\mathrm{sky}} \frac{\delta_{ij}}{N^X(x_i)} \, , 
    \label{eq:noise_resolved}
\end{equation}
where $f^X_{\mathrm{sky}}$ is the fraction of the sky covered by the observation of tracer $X$, $N^X(x_i)$ represents the number of events of tracer $X$ in the $i$--th redshift bin, and $\delta_{ij}$ is the Kronecker delta.
In addition to noise, to completely describe the angular power spectrum, it is necessary to account for the damping of the signal due to the finite localization capabilities of the detectors.
Assuming a circular Gaussian profile, the beam window function which accounts for the localization errors for a given tracer $X$ at the $i$--th redshift bin is defined as
\begin{equation}
    \mathcal{B}^X_\ell(x_i) = \exp \biggl\{ - \frac{ \ell (\ell+1) [ \sigma^X(x_i) ] ^2}{2} \biggr\} \, ,
\end{equation}
where $\sigma^X(x_i)$ is the beam width of tracer $X$ at the $i$--th redshift bin.
In contrast, as mentioned previously, \ac{hi} \ac{im} measurements do not resolve each source individually, but instead observe the overall emission, thus not allowing for the application of Equation~\eqref{eq:noise_resolved}.
In the case of unresolved sources, the formalism for the derivation of noise equations is tracer-dependent.
The specific equations for the two tracers relevant to our analysis will be presented in section~\ref{sec:tracers}.

\subsection{Gravitational Waves and Neutral Hydrogen as large-scale structure tracers}\label{sec:tracers}
The topic of cross-correlation between \ac{gw} sources and \ac{lss} tracers has gained significant attention in recent years.
Past works, such as Ref.~\cite{Scelfo:2021fqe}, demonstrated the theoretical viability of using \ac{hi} \ac{im} as a tracer.
However, these early analyses were limited to Fisher matrix forecasts for cosmological parameters, which assume purely Gaussian posteriors, resulting in a lower limit on variance and a limited resolution on the degeneracy between the parameters.
More recently, full Bayesian \ac{mcmc} pipelines have been used to overcome these statistical limitations.
For instance, Refs.~\cite{sala2025, pedrotti2025} successfully extracted cosmological constraints through full \ac{mcmc} analyses of the cross-correlation signal between \ac{gw} dark sirens and photometric galaxy surveys.

In this work, we go beyond the analyses of Ref.~\cite{Scelfo:2021fqe} by applying the full Bayesian \ac{mcmc} machinery to the synergy between next-generation \ac{gw} observatories (\ac{et}) and 21\,cm \ac{hi} \ac{im} (\ac{skao}).
To move beyond the idealized framework of Ref.~\cite{Scelfo:2021fqe} we introduce several improvements to the analysis:
\begin{itemize}
    \item we use \texttt{GWFish+Priors}~\cite{DupletsaHarms2023, Dupletsa:2024gfl} to simulate realistic \ac{gw} mock catalogs, which allows us to account for parameter prior bounds into the Fisher approximation. 
    This replaces the simplified fixed-error approach, providing event-specific uncertainty estimates on key source properties such as the luminosity distance, $d_L$, and sky localization, $\Delta\Omega$, which depend on redshift, source properties and detector sensitivity; 
    \item we choose a tomographic binning strategy that approximately puts the same number of sources in each bin, thus ensuring that each bin has a similar statistical weight across the entire redshift range;
    \item as discussed previously, we compute the exact, beyond-Limber fully relativistic angular power spectra.
\end{itemize}

We focus the analysis on the redshift range $0.5 \le z \le 3.5$, which corresponds to the maximal overlapping redshift range between \ac{skao} and next-generation \ac{gw} detectors like \ac{et}.
Including \ac{gw} events and \ac{hi} signal at higher or lower redshifts would not significantly improve the analysis, since, due to instrumental limitations and, especially, the low \ac{gw} event rates, the correspondence between the two tracers would be reduced, thus resulting in a negligible cross-correlation signal~\cite{Scelfo:2021fqe}.

Throughout this work, we assume the standard $\Lambda$ \ac{cdm} cosmological model.
This model assumes a cosmological constant ($\Lambda$) driving the accelerated background expansion and cold, collisionless \ac{dm} driving the formation of \ac{lss}.
In this framework, the background expansion is parameterized by the Hubble constant, $H_\mathrm{0}$, and the physical matter density, $\omega_ \mathrm{cdm}$.
At the level of linear perturbations, the growth of cosmic structures, which drives the spatial clustering of both our tracers (\ac{gw} and \ac{hi}), is described by the amplitude and tilt of the primordial power spectrum: $\ln(10^{10}A_\mathrm{s})$ and $n_\mathrm{s}$, respectively.

\begin{figure}[ht]
    \centering
    \begin{subfigure}[b]{0.49\columnwidth}
        \centering
        \includegraphics[width=\linewidth]{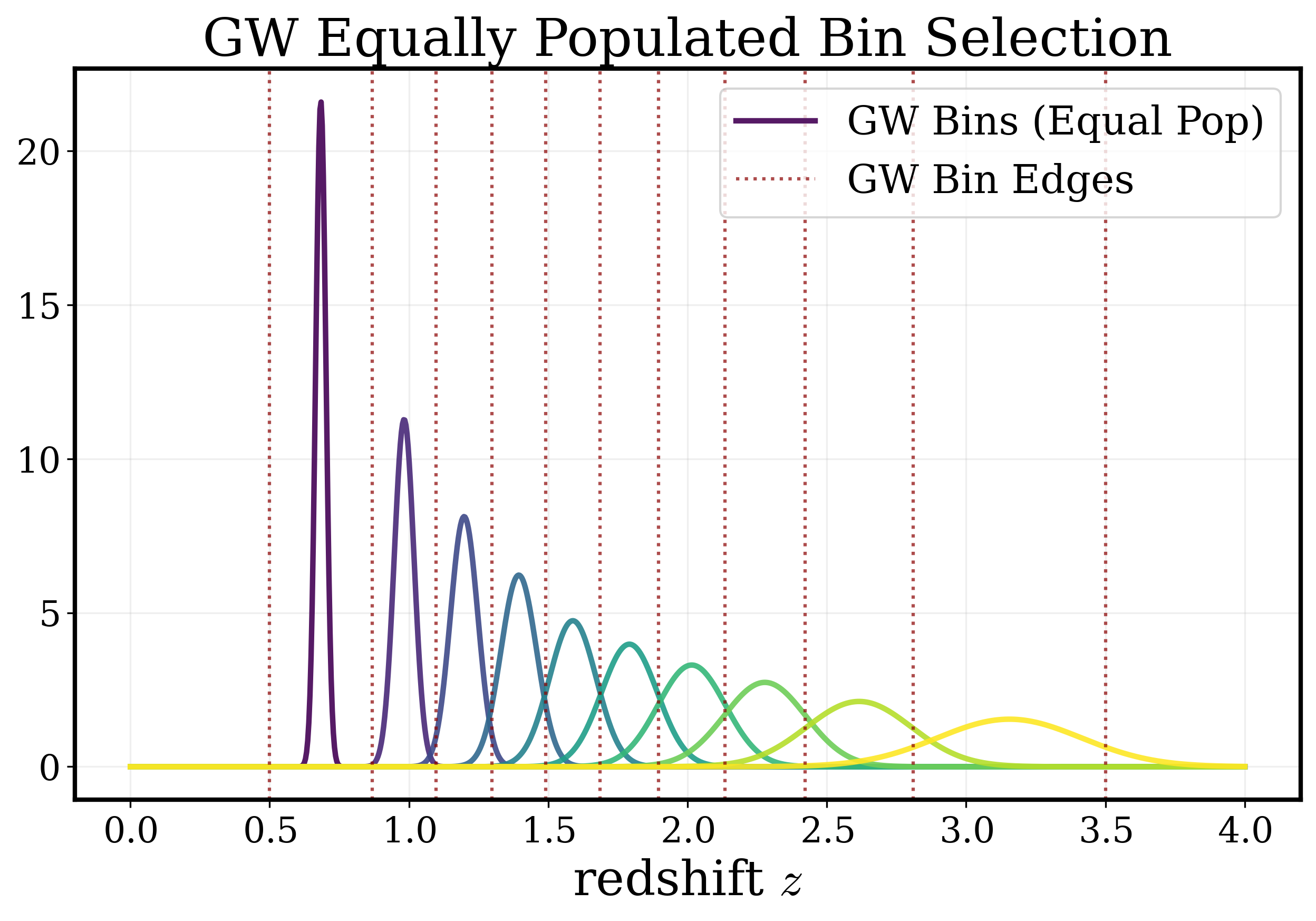} 
        \caption{\ac{gw} binning}
        \label{fig:bins_fixed}
    \end{subfigure}
    \hfill
    \begin{subfigure}[b]{0.49\columnwidth}
        \centering
        \includegraphics[width=\linewidth]{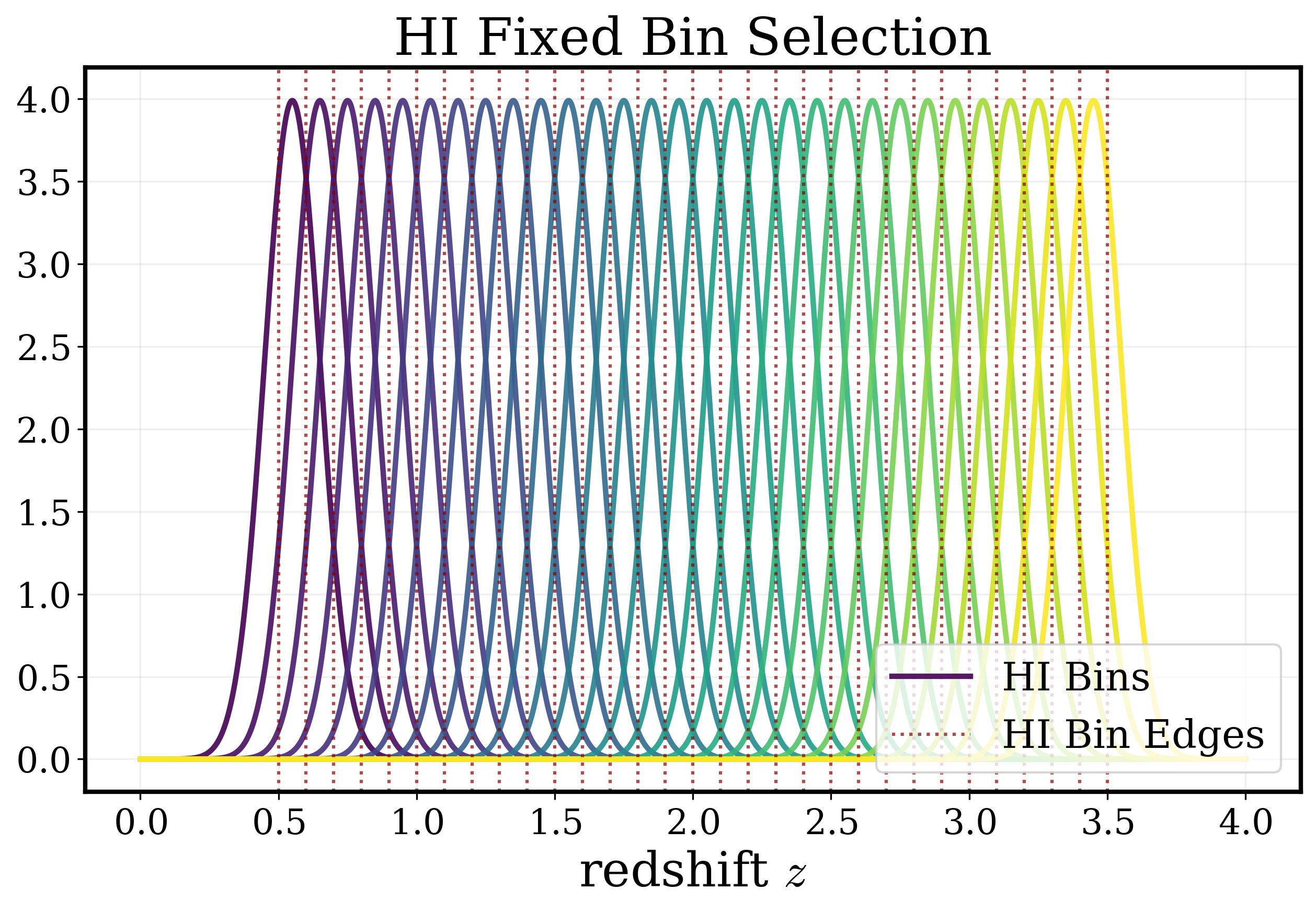} 
        \caption{\ac{hi} binning}
        \label{fig:bins_equal}
    \end{subfigure}
    \caption{Redshift binning strategy for \ac{gw}s (a) and for \ac{hi} (b). For \ac{gw}s, the bin widths shown correspond to the ET2L+CE detector configuration, each containing approximately the same number of sources ($N_{src} \approx \text{const}$). The color code has no physical meaning, it is only used for a better visualization of the different bins.}
    \label{fig:bin_choices}
\end{figure}

As illustrated in Figure~\ref{fig:bin_choices}, we adopt two different choices for the redshift bins of the two tracers:
\begin{itemize}
    \item \textit{Fixed bins for \ac{hi}:} narrow redshift bins for \ac{hi} $\Delta z_{\textrm{H\textsc{i}}} = 0.1$, following the binning strategy used in~\cite{Scelfo:2021fqe}.
    This is expected to be the optimal redshift range for \ac{skao}~\cite{Bacon_Battye_2020}, and the high precision on redshift of spectroscopic surveys allows for fine tomographic bins with negligible radial uncertainties.
    
    \item \textit{Equally populated bins for \ac{gw}:} redshift bin widths are defined based on the simulated tracer populations fixing a fiducial cosmology.
    Instead of fixing a constant $\Delta z$, the bin edges are adjusted so that each tomographic slice contains approximately the same number of sources, as described in \cite{sala2025}.
    By doing so, we ensure that the statistical weight of each bin is similar across the entire redshift range.
    Since \ac{gw} detection rates decrease at high redshifts, high-$z$ bins are widened to enclose the target number of sources.
    This prevents high-redshift tomographic slices from being dominated by shot noise, ensuring a uniform level of statistical information across the entire redshift range.
\end{itemize}

\subsubsection{Gravitational Waves from Binary Black Holes}\label{sec:gw_description}
We will perform our analyses focusing on \ac{gw} signals from \ac{bbh} mergers, as they would be observed by three different \ac{gw} networks~\cite{Branchesi_2023}: \textit{(i)} ET in its $\Delta$ configuration (ET-$\Delta$) located in Sardinia; \textit{(ii)} \ac{et} in its 2L configuration (ET-2L), one located in Sardinia and one in Saxony; and \textit{(iii)} ET-2L together with one Cosmic Explorer (CE) of 40 km located in the US (ET2L+CE).
Building upon Refs.~\cite{Scelfo:2021fqe, pedrotti2025, sala2025}, we adopt the prescriptions highlighted below.

\begin{description}
    \item[\ac{bbh} redshift distribution:] we evaluate the expected population of \ac{gw} events by generating a realistic population of \ac{bbh} mergers.
    To model the intrinsic properties of the binaries, we sample the mass distributions from a fiducial ``PowerLaw+Peak'' distribution and draw the redshift distribution according to the Madau-Dickinson star formation history, consistently with the results from \cite{Abbott_2023}, by means of \texttt{icarogw}~\cite{mastrogiovanni2023icarogwpythonpackageinference}.
    From the resulting mock population, we compute the distribution of \ac{bbh} observed by different \ac{gw} detector networks using \texttt{GWFish+Priors}~\cite{DupletsaHarms2023,Dupletsa:2024gfl}.
    Note that the inclusion of physical priors naturally skews the standard Gaussian uncertainties obtained with Fisher analyses.
    To avoid losing statistical information by modeling the errors with a simple Gaussian standard deviation ($\sigma$), we employ Kernel Density Estimation (KDE)~\cite{Silverman1986, Scott1992}.
    By doing so, we take into account the full non-Gaussian posterior distributions of the parameters, ensuring our analysis relies on the exact shape of their errors rather than just their variance.
    Using the \texttt{IMRPhenomXHM} waveform model and accounting for the detector response and duty cycle, we isolate the detectable events by applying a signal-to-noise ratio threshold of $\rho \geq 8$.
    The resulting redshift distribution profile, $dN_\mathrm{GW}^\mathrm{det}/dz$ including both the underlying astrophysical merger rate and the network-dependent detection efficiencies reads~\cite{abbott2023population, mastrogiovanni2023icarogwpythonpackageinference}:
    \begin{equation}
        \frac{dN_\mathrm{GW}^\mathrm{det}}{dz} = \alpha_\mathrm{GW} z^{\beta_\mathrm{GW}} \exp{ \left[ - \gamma_\mathrm{GW} z \right] } \, ,
        \label{eq:gw_dndz_scelfo}
    \end{equation}
    where the parameters strictly depend on the network considered in the analysis.
    The values corresponding to the different networks are reported in Table~\ref{tab:dndz_parameters}. They are obtained by assuming full sky coverage and 1\,year of observations ($f_\mathrm{sky}=1.0$ and $t_\mathrm{obs}^\mathrm{GW} = 1\mathrm{yr}$).
    The total number of \ac{bbh} mergers detections within the redshift range $0.5 \leq z \leq 3.5$ is $\sim$\,$10^5$ per year\footnote{Since $f_\mathrm{sky}$ and $t_\mathrm{obs}^\mathrm{GW}$ act like scaling factors, different configurations can be easily examined by just re-scaling the overall distribution.}.
    
    \begin{table}[htpb]
        \centering
        \begin{tabular}{lccc}
            \toprule
            \textbf{Parameter} & \textbf{ET-$\Delta$} & \textbf{ET-2L} & \textbf{ET2L+CE} \\
            \midrule
            $\alpha_\mathrm{GW}$ & $4.129 \times 10^5$ & $3.633 \times 10^5$ & $3.243 \times 10^5$ \\
            $\beta_\mathrm{GW}$ & 3.2763 & 3.1699 & 3.0532 \\
            $\gamma_\mathrm{GW}$ & 2.4360 & 2.3106 & 2.1555 \\
            $N_\mathrm{GW}^\mathrm{det}$ (1\,yr) & $\sim 9.6\times 10^4$ & $\sim 8.6\times 10^4$ & $\sim 8.2\times 10^4$ \\
            \bottomrule
        \end{tabular}
        \caption{Best-fit parameters for the phenomenological redshift distribution of the detected \ac{bbh} events for the ET-$\Delta$, ET-2L, and ET2L+CE networks. The normalized probability density function is modeled as $\frac{dN_\mathrm{GW}^\mathrm{det}}{dz}  = \alpha_\mathrm{GW} \cdot z^{\beta_\mathrm{GW}}\cdot e^{-\gamma_\mathrm{GW} z}$.
        For each respective detector configuration, $N_\mathrm{GW}^\mathrm{det}$ denotes the total number of expected \ac{bbh} detections per year.}
        \label{tab:dndz_parameters}
    \end{table}

    \item[Biases:] three different bias terms have to be included to correctly describe the \ac{gw} distribution with respect to the underlying \ac{dm} field:
    \begin{itemize}
        \item the \textit{selection bias}, which describes the relation between a given observable $X$ and the underlying distribution of matter that it traces.
        Assuming a linear bias model, it is defined as $b_X(x_i) = \delta_X(x) / \delta_{DM}(x)$ and takes the following phenomenological form~\cite{Scelfo:2020jyw}:
            \begin{equation}
                b_\mathrm{GW}(z) = a_\mathrm{GW} \exp{ \left[ b_\mathrm{GW}z^{d_\mathrm{GW}} \right]} + z^{c_\mathrm{GW}}
                \label{eq:gw_bias};
            \end{equation}
            with $a_\mathrm{GW} = 0.948$, $b_\mathrm{GW} = -0.553$, $c_\mathrm{GW} = 0.996$, and $d_\mathrm{GW} = 1.034$~\cite{Libanore_2021};
        \item the \textit{magnification bias}, which quantifies how the observed surface density of sources of tracer $X$ is influenced by gravitational lensing effects~\cite{turner_1984}.
        It is defined as the logarithmic slope of the redshift distribution of the detected events computed at detection limit $\rho = \rho_\mathrm{thr}$:
            \begin{equation}
                s_\mathrm{GW}(z) = - \left. \frac{d \log{ \left( \frac{d^2 N_\mathrm{GW}(z,>\rho_\mathrm{thr})}{dzd\Omega} \right) }}{d\rho} \right|_{\rho = \rho_\mathrm{thr}}
                \label{eq:gw_magnbias} \, ;
            \end{equation}
            
        \item the \textit{evolution bias}, which encodes the evolution over time of the absolute number of sources of the tracer $X$.
        It is defined as:
            \begin{equation}
                f_\mathrm{GW}^\mathrm{evo}(z) = \frac{d \ln{ \left( a^3 \frac{d^2 N_\mathrm{GW}(z)}{dzd\Omega} \right) }}{d \ln{a}} \, ,
                \label{eq:gw_evobias}
            \end{equation}
            where $a$ is the cosmological scale factor, and $\frac{d^2 N_\mathrm{GW}(z)}{dzd\Omega}$ is the redshift distribution of resolved sources.
    \end{itemize}
    
    \item[Noise:] as already discussed, for resolved sources such as \ac{gw}s, the main contribution to the noise is the shot noise, described by Equation~\eqref{eq:noise_resolved}.
    Additionally, the finite angular resolution of the \ac{gw} network produces an effective damping of the signal at small scales, which is given by~\cite{Oguri_2016, Libanore_2021, Scelfo_2023}
    \begin{equation}
        \mathcal{B}^\mathrm{GW}(z_j) = \exp{\left[ -\frac{\ell (\ell+1)}{\ell_\mathrm{damp}^2} \right]} \, ,
        \label{eq:gw_beaming}
    \end{equation}
    \begin{equation}
        \ell_\mathrm{damp}^2 = \frac{ (2 \pi)^{3/2} }{\Delta \Omega_\mathrm{1\sigma}}
        \label{eq:ell2_damp} \, ,
    \end{equation}
    where $\Delta \Omega_\mathrm{1\sigma}$ is the solid angle corresponding to the 68\% contour level of the angular error of the \ac{gw} network.
    The specific values for each \ac{gw} detector network configuration considered in this work are reported in Table~\ref{tab:analytic_z_bins}.
    We compute these effective sky-localization capabilities directly from our mock \ac{bbh} catalogs generated via \texttt{GWFish+Priors}~\cite{DupletsaHarms2023, Dupletsa:2024gfl}. 
    The resulting values reflect the angular resolution of the different configurations of the next-generation \ac{gw} detectors~\cite{Maggiore_2020}.
    Assuming $f_\mathrm{sky}=1.0$, in the auto-correlation case the simulated angular power spectra will read:
    \begin{equation}
        \begin{split}
            \tilde{C_\ell}^\mathrm{GW,GW}(z_i, z_j) &= \mathcal{B}^\mathrm{GW}(z_i) \mathcal{B}^\mathrm{GW}(z_j) C_{\ell,th}^\mathrm{GW,GW}(z_i, z_j) + \mathcal{N}_\ell^\mathrm{GW}(z_i,z_j) \\
            &= \mathcal{B}^\mathrm{GW}(z_i) \mathcal{B}^\mathrm{GW}(z_j) C_{\ell,th}^\mathrm{GW,GW}(z_i, z_j) + \frac{\delta_{ij}}{\bar{N}^\mathrm{GW}(z_i)} \, ,
        \end{split}
        \label{eq:gwgw_observedCls}
    \end{equation}
    where $\bar{N}^\mathrm{GW}(z_i)$ is mean number density of \ac{gw}s in the $i$-th redshift bin.
    Notably, the shot noise affects only the same-bin auto-correlation components of the angular power spectra.
    
    \begin{table}[htpb]
        \centering
        \begin{tabular}{lccccc}
            \toprule
             & ET-$\Delta$ & ET-2L & ET2L+CE \\
            \textbf{Bin Edges} & $\ell_{\mathrm{damp}}$ & $\ell_{\mathrm{damp}}$ & $\ell_{\mathrm{damp}}$ \\
            \midrule
            $0.500 - 0.867$ & 7 & 20 & 229 \\
            $0.867 - 1.098$ & 6 & 16 & 171 \\
            $1.098 - 1.298$ & 7 & 16 & 149 \\
            $1.298 - 1.490$ & 6 & 15 & 131 \\
            $1.490 - 1.686$ & 7 & 14 & 114 \\
            $1.686 - 1.896$ & 6 & 13 & 106 \\
            $1.896 - 2.134$ & 6 & 13 & 95  \\
            $2.134 - 2.421$ & 6 & 13 & 87  \\
            $2.421 - 2.810$ & 6 & 12 & 78  \\
            $2.810 - 3.500$ & 6 & 12 & 69  \\
            \bottomrule
        \end{tabular}
        \caption{Tomographic redshift bins derived from the analytic phenomenological distribution. For each bin, we report the redshift range and the cross-correlation damping scales $\ell_{\mathrm{damp}}$ for the ET-$\Delta$, ET-2L, and ET2-CE networks.}
        \label{tab:analytic_z_bins}
    \end{table}    
\end{description}

\subsubsection{Neutral Hydrogen Intensity Mapping} \label{sec:hi}
We simulate the second tracer, \ac{hi} \ac{im}, assuming an ideal 21\,cm emission single-dish survey with the upcoming \ac{skao}~\cite{Bacon_Battye_2020, santos2015cosmologyskahiintensity}.

Due to cosmic expansion, the rest-frame 21\,cm emission of \ac{hi} is redshifted to lower frequencies, $\nu_{\mathrm{obs}}$, according to:
\begin{equation}\label{eq:hi_rest_freq}
    \nu_{\mathrm{obs}} = \nu_\mathrm{21cm}/(1+z) \,,~~ {\rm with }~~\nu_\mathrm{21cm}=1420~{\rm MHz} \, .
\end{equation}
Observing below the GHz band thus opens up a valuable observational window into the past of the Universe.
In this analysis, we focus on SKA-Mid operating in band\,2 ($950-1760$\,MHz) and band\,1 ($350-1050$\,MHz), which can cover the redshift range $0 \lesssim z \lesssim 3$.
To comprehensively map the \ac{hi} density distribution throughout the entire post-reionization Universe ($z \lesssim 6$), we additionally assume the availability of \ac{hi} \ac{im} data from SKA-Low ($50-350$\,MHz)\footnote{During the reionization era, the underlying astrophysics becomes significantly more complex. Consequently, the \ac{hi} field cannot be treated simply as a standard \ac{lss} tracer, and its interpretation requires modeling additional astrophysical effects, e.g.~\cite{greig_2017}.}.

The key idea behind \ac{im} is to exploit the collective 21\,cm emission from many unresolved objects, rather than detect individually resolved sources, since the faint 21\,cm line would otherwise limit detections to the local Universe.
This approach enables mapping the \ac{lss} at lower angular resolution but to unprecedented depths (e.g. ~\cite{battye_2013, kovetz2017lineintensitymapping2017status}).

Tracing the distribution of \ac{hi}, \ac{im} surveys conducted with radio telescopes can probe the underlying \ac{dm} field on large scales out to the highest redshifts allowed by their frequency coverage.
In particular, we assume that with SKA-Mid in single-dish mode we can recover the large-scale clustering signal up to $z \sim 3$, while higher redshift can be covered with SKA-Low~\cite{SKA:2018ckk}. 

As the traditional angular power spectrum formalism applies only to resolved sources, when applying this framework to the (unresolved) \ac{hi} \ac{im} signal, some specific adaptations have to be taken into account~\cite{Hall_2013, Alonso:2015sfa}.
Our prescriptions are outlined below.

\begin{description}
    \item[Redshift distribution:] instead of considering the number count redshift distribution $dN_X/dz$, the \ac{hi} redshift distribution can be described via its comoving density distribution $\rho_{\textrm{H\textsc{i}}}(z)$ and its mean brightness temperature distribution $T_b(z)$~\cite{crighton_2015, battye_2013}:
    \begin{equation}
        \rho_\mathrm{\textrm{H\textsc{i}}}(z) = \Omega_\mathrm{\textrm{H\textsc{i}}}(z) \rho_\mathrm{crit,0} = 4 (1+z)^{0.6} 10^{-4} \rho_\mathrm{crit,0} \, ,
    \end{equation}
    \begin{equation}
        T_b(z) = 44\mu K \left( \frac{\Omega_\mathrm{\textrm{H\textsc{i}}}(z)h}{2.45 \times 10^{-4}} \right) \frac{(1+z)^2}{E(z)}\, ,
    \end{equation}
    where $\rho_\mathrm{crit,0}$ is the critical density today, $\Omega_\mathrm{\textrm{H\textsc{i}}}(z) = \rho_\mathrm{\textrm{H\textsc{i}}}(z)/\rho_\mathrm{crit,0}$ is the hydrogen density abundance distribution, $h = H_0/(100~\mathrm{km~s^{-1}~Mpc^{-1})} $ is the dimensionless Hubble parameter, and $E(z) = H(z)/H_0$ describes the expansion rate of the Universe.
    In particular, the mean brightness temperature $T_b$ is the quantity directly observed by the observatory and therefore describes the effective observed redshift distribution\footnote{The brightness temperature serves as a direct tracer of the underlying density because the 21\,cm emission is optically thin on cosmological scales.
    In this regime, the observed intensity is linearly proportional to the number density of neutral hydrogen atoms.} which has to be included in Equation~\eqref{eq:unnorm_windowfunc}.
    By contrast, the comoving density distribution $\rho_\mathrm{\textrm{H\textsc{i}}}$ represents the redshift dependence of the absolute mass density of \ac{hi}, thus entering only through the evolution bias term (see Equation~\eqref{eq:hi_evobias}).

    \item[Biases:] similar to the case of resolved sources, three different bias terms have to be included to correctly describe the \ac{hi} component with respect to the underlying \ac{dm} distribution:
    \begin{itemize}
        \item \textit{selection bias}, which takes the following phenomenological form~\cite{spinelli_2020}:
            \begin{equation}
                b_\mathrm{\textrm{H\textsc{i}}}(z) = a_\mathrm{\textrm{H\textsc{i}}} (1+z)^{b_\mathrm{\textrm{H\textsc{i}}}} + c_\mathrm{\textrm{H\textsc{i}}}
                \label{eq:hi_bias}\, ,
            \end{equation}
            where $a_\mathrm{\textrm{H\textsc{i}}} = 0.22$, $b_\mathrm{\textrm{H\textsc{i}}} = 1.47$, and $c_\mathrm{\textrm{H\textsc{i}}} = 0.63$ are  parameters obtained by fitting results from Ref.~\cite{spinelli_2020}.
            Equation~\eqref{eq:hi_bias} is derived from a semi-analytical galaxy formation model that accounts for neutral hydrogen.
            The results are consistent with findings of \cite{Villaescusa-Navarro_2018} based on the Illustris TNG hydrodynamical simulations;
        \item \textit{magnification bias}: as discussed in \cite{Hall_2013}, when considering \ac{im} experiments lensing effects are absent, corresponding to a constant value for the magnification bias, namely
            \begin{equation}
                s_\mathrm{\textrm{H\textsc{i}}}(z) = 0.4 \, ;
                \label{eq:hi_magnbias}
            \end{equation}
        \item \textit{evolution bias}, defined as:
            \begin{equation}
                f_\mathrm{\textrm{H\textsc{i}}}^\mathrm{evo}(z) = \frac{d \ln{\rho_\mathrm{\textrm{H\textsc{i}}}(z)}}{d \ln{a}}\, ,
                \label{eq:hi_evobias}
            \end{equation}
            where $a$ is the cosmological scale factor, and $\rho_{HI}(z)$ is the \ac{hi} density distribution.
    \end{itemize}

    \item[Noise:] in the specific case of \ac{hi}, the simulated angular power spectrum is affected by both observational noise and finite beam size of the instrument.
    The correlation signal is suppressed for scales smaller than the Full Width at Half Maximum (FWHM) of the beam $\theta_B(z) = 1.22 \lambda_{21\mathrm{cm}}(1+z) / D_d$, where $D_d$ is the diameter of a single dish of the interferometer, and is therefore given by:
    \begin{equation}
        \mathcal{B}^\mathrm{\textrm{H\textsc{i}}}(z_i) = \exp{\left[ -\ell (\ell+1) \left( \frac{\theta_B(z_i)}{\sqrt{16\ln{2}}} \right)^2 \right]} \, .
        \label{eq:hi_suppression}
    \end{equation}
    The simulated \ac{hi} auto-correlation spectrum is given by
    \begin{equation}
        \tilde{C_\ell}^\textrm{H\textsc{i},H\textsc{i}}(z_i, z_j) = \mathcal{B}^\textrm{H\textsc{i}}(z_i) \mathcal{B}^\textrm{H\textsc{i}}(z_j) C_{\ell,th}^\textrm{H\textsc{i},H\textsc{i}}(z_i, z_j) + \delta_{ij} \mathcal{N}_\ell^\textrm{H\textsc{i}}(z_i) \, ,
        \label{eq:hihi_observedCls}
    \end{equation}
    while the cross-correlation term is given by
    \begin{equation}
        \tilde{C_\ell}^\textrm{H\textsc{i},GW}(z_i, z_j) = \mathcal{B}^\mathrm{\textrm{H\textsc{i}}}(z_i) \mathcal{B}^\mathrm{GW}(z_j) C_{\ell,th}^\mathrm{\textrm{H\textsc{i}},GW}(z_i, z_j), 
        \label{eq:higw_observedCls}
    \end{equation}
    where $\mathcal{B}^\mathrm{GW}(z_j)$ is defined in Equation~\eqref{eq:gw_beaming}, and
    \begin{equation}
        \mathcal{N}_\ell^\mathrm{\textrm{H\textsc{i}}}(z_i) = \mathcal{N}_\ell^\mathrm{instr}(z_i) + \mathcal{N}_\ell^\mathrm{fg}
        \label{eq:hi_noise}
    \end{equation}
    represents the total noise spectrum for \ac{hi}.
    Specifically, $\mathcal{N}_\ell^\mathrm{instr}(z_i)$ accounts for the intrinsic noise of the instrument and $\mathcal{N}_\ell^\mathrm{fg}$ describes the residual error due to the procedure of cleaning the cosmological \ac{im} signal from the bright foreground emission.
    We parametrize these two noise components as described below.
    \begin{itemize}
        \item \textit{Instrumental noise.} It is strictly linked to the observational setup.
        We consider \ac{skao} in its single-dish configuration, with a collection of $N_\mathrm{dish} = 254$ dishes. This leads to the following expression for the instrumental component of the noise angular spectrum~\cite{santos2017meerklassmeerkatlargearea, santos2015cosmologyskahiintensity, Bull_2015}:
        \begin{equation}
            \mathcal{N}_\ell^\mathrm{instr}(z_i) = \sigma_T^2(z_i) \theta_B^2 \approx \left( \frac{T_\mathrm{sys}}{T_b(z_i) \sqrt{n_\mathrm{pol} B t_\mathrm{obs}} } \sqrt{\frac{S_\mathrm{area}}{\theta_B^2}} \right)^2 \theta_B^2 \, ,
            \label{eq:instrm_noise}
        \end{equation}
        where $\sigma_T(z_i)$ is the single-dish root mean square noise temperature, $T_\mathrm{sys} = 28 \mathrm{K}$ for the system temperature, $B = 20 \cdot 10^6 ~\mathrm{Hz}$ for the bandwidth, $t_\mathrm{obs} = 1.8 \cdot 10^7 \mathrm{s}$ as the observation time, and $S_\mathrm{area} = 20000 ~\mathrm{deg}^2$ for the total surveyed area (corresponding to $f_\mathrm{sky} = 0.5$)
        Since the instrumental noise is inherently independent across different observational frequencies, we assume it to be completely uncorrelated between distinct tomographic slices.
        For this reason, the instrumental noise component affects only the auto-correlation spectrum of a single redshift bin ($z_i = z_j$), while it vanishes for all cross-correlation terms.
        
        \item \textit{Foreground}. Thanks to the increased sensitivity and spectral coverage of \ac{skao}, the strong foregrounds that affect and prevent the detection of the \ac{im} auto-correlation signal in current observations~\cite{switzer_2013, Wolz_2021} are expected to be overcome. 
        Nevertheless, imperfect cleaning, due to the complexity of the foregrounds, will leave residual foreground emission in angular power spectra~\cite{alonso_2014, Carucci_2020, Soares_2021, Matshawule_2021, Cunnington_2021}.
        To account for this post-removal error, we add the noise term $\mathcal{N}_\ell^\mathrm{fg}$ to the theoretical angular power spectra for all \ac{im} components, including cross-bin auto-correlation as shown in Equation~\eqref{eq:hihi_observedCls}.
        The residual noise takes the following form:
        \begin{equation}
            \mathcal{N}_\ell^\mathrm{fg} = K^\mathrm{fg} \cdot F(\ell) =  K^\mathrm{fg} \frac{1}{f_\mathrm{sky}} A_\mathrm{fg} e^{b_\mathrm{fg} \ell^{c_\mathrm{fg}}} \, .
            \label{eq:fg_noise}
        \end{equation}
        Here $K^\mathrm{fg} \simeq 6 \cdot 10^{-7}$ is a normalization constant derived from the average value of $C_\ell^\mathrm{\textrm{H\textsc{i}},\textrm{H\textsc{i}}}(z_i,z_j)$, and $F(\ell)$ describes the scale-dependency of the residual foregrounds noise, where, based on results of \cite{alonso_2014}, the single parameters are: $A_\mathrm{fg} \sim 0.129$, $b_\mathrm{fg} \sim -0.081$, and $c_\mathrm{fg} \sim 0.581$.
    \end{itemize}
    
\end{description}

\subsection{Cosmological parameter estimation}\label{sec:likelihood}
To build our likelihood pipeline, we have extended the statistical framework applied in Refs.~\cite{sala2025, dematos2025} to galaxy catalogs, adapting it to the case of $\ac{gw}\times\ac{hi}$ cross-correlation.
To correctly describe the $C_\ell$--based likelihood it is crucial to organize both the simulated ($\tilde{C}_\ell(x_i,x_j)$) and the theoretical ($C_\ell^{th}(x_i, x_j, \Gamma)$) angular power spectra\footnote{We will refer with $\sim$ simulated spectra and with $th$ theoretical spectra.} to have the following structure:
\begin{equation}
    \mathcal{C}_\ell =
    \begin{pmatrix}
        \mathbf{C_\ell^{XX}} \\
        \mathbf{C_\ell^{XY}}\\
        \mathbf{C_\ell^{YY}}
    \end{pmatrix} \, ,
\end{equation}
where $\mathbf{C_\ell^{XX}} = C_\ell^{XX}(x_j \ge x_i)$ is the sub-vector containing all non-trivial auto-correlations and cross-correlations among all the different redshift bins for tracer $X$.
Similarly, $\mathbf{C_\ell^{YY}}$ is defined considering tracer $Y$.
Differently, $\mathbf{C_\ell^{XY}} = C_\ell^{XX}(x_i, x_j) $ is the sub-vector containing all auto-correlations and cross-correlations among all the different redshift bins for tracers $X$ and $Y$.
Therefore, if we consider $n$ redshift bins for tracer $X$ and $m$ for tracer $Y$, then, \textit{for a given multipole $\ell$}, the $XX$ vector will contain $\left[ n ( n +1) \right] /2$ elements (similarly vector $YY$ will contain $\left[ m ( m +1) \right] /2$), while vector $XY$ will have $\left[ (n +m) ( n+m+1) \right] /2$ entries\footnote{Note that in the cross-correlation term $C_\ell^{XY}(x_1,x_2) \neq C_\ell^{XY}(x_2,x_1)$}.
For clarity, we report the explicit form of $\mathcal{C_\ell}$:
\begin{equation}
    \mathcal{C}_\ell =
    \begin{pmatrix}
        \left( C_\ell^{XX}(x_1,x_1), C_\ell^{XX}(x_1,x_2), ..., C_\ell^{XX}(x_{n},x_{n}) \right)^T, \\ 
        \left( C_\ell^{XY}(x_1,x_1), C_\ell^{XY}(x_1,x_2), ..., C_\ell^{XY}(x_2,x_1), C_\ell^{XY}(x_2,x_2), ..., C_\ell^{XY}(x_{n},x_{m}) \right)^T, \\
        \left( C_\ell^{YY}(x_1,x_1), C_\ell^{YY}(x_1,x_2), ..., C_\ell^{YY}(x_{m},x_{m}) \right)^T
    \end{pmatrix} \, .
    \label{eq:cl_matrix}
\end{equation}
Given the definition in Equation~\eqref{eq:cl_matrix}, the explicit expression for the $C_\ell$--based likelihood reads:
\begin{equation}
    \ln \mathcal{L} (\Gamma) = \sum_{\ell = \ell_{min}}^{\ell_{max}}  \sum_{ij}  \sum_{i'j'}  \Delta \mathcal{C}_\ell^{ij} \left[ \mathrm{Cov}_\ell^{-1} \right]\Delta \mathcal{C}_\ell^{i'j'} \, ,
    \label{eq:likelihood}
\end{equation}
where $\Gamma$ denotes the set of cosmological parameters on which the inference is carried out, $\ell_{min}$ and $\ell_{max}$ are the minimum and maximum multipoles considered in the analysis, and are linked to the observational capabilities of the detectors.
The quantity
\begin{equation}
    \Delta \mathcal{C}_\ell^{ij} = \tilde{\mathcal{C}}_\ell(x_i,x_j) - \mathcal{C}_\ell^{th}(x_i, x_j, \Gamma)
\end{equation}
describes the discrepancy between the simulated spectra and the theoretical prediction for a given set of cosmological parameters $\Gamma$.
The covariance matrix for a given multipole is given by
\begin{equation}
    \mathrm{Cov}_\ell^{-1} = \mathrm{Cov} \left[ C_{\ell, ij}^{IJ}, C_{\ell', i'j'}^{I'J'} \right] 
    = \frac{\delta_{\ell\ell'}}{(2\ell +1) f_\mathrm{sky}} \left[ 
    \tilde{C}_{\ell, ii'}^{II'} \tilde{C}_{\ell, jj'}^{JJ'} + \tilde{C}_{\ell, ij'}^{IJ'} \tilde{C}_{\ell, ji'}^{JI'} \right] \, ,
\end{equation}
where $f_\mathrm{sky}$ is the fraction of the sky covered by the observational surveys, $I$ and $J$ are the indices of the tracers $X$ and $Y$, and $i$ and $j$ run over the redshift bins.
Note that the covariance matrix has to be inverted for each separate multipole $\ell$ independently, before being summed to compute the full likelihood.

Additionally, since we assume linear perturbation theory, there is no correlation between different multipoles, as perturbation at different scales are independent.

\begin{table}[ht]
    \centering
    \renewcommand{\arraystretch}{1.2}
    \begin{tabular}{@{}l c c@{}} 
        \toprule
        \textbf{Parameter} & \textbf{Fiducial Value} & \textbf{Prior} \\
        \midrule
        \midrule
        $\omega_\mathrm{cdm}$ & $0.1188$ & $\mathcal{U}[0.001, 0.99]$ \\
        $H_\mathrm{0}$ [km s$^{-1}$ Mpc$^{-1}$] & $67.74$ & $\mathcal{U}[30, 100]$ \\
        $\ln(10^{10}A_\mathrm{s})$ & $3.04478$ & $\mathcal{U}[2, 4]$ \\
        $\omega_b$ & $0.02230$ & -- \\
        $\tau_\mathrm{reio}$ & $0.0544$ & -- \\
        $n_s$ & $0.96605$ & -- \\
        \midrule
        $\Omega_\mathrm{m}$ & $0.3085$ & -- \\
        $\sigma_8$ & $0.8228$ &-- \\        
        \bottomrule
    \end{tabular}
    \caption{Fiducial values and prior distributions for the cosmological parameters considered in the analysis. The fiducial (injected) values are set according to \textit{Planck} results \cite{Ade2015PlanckXIII, Aghanim2018PlanckVI}. The priors for the sampled parameters are assumed to be uniform ($\mathcal{U}$) within the specified ranges. The parameters derived at each MCMC step ($\Omega_\mathrm{m}$, $\sigma_8$) and the background $\Lambda$CDM parameters held fixed during the analysis are also reported as rows with no value in the prior column.}
    \label{tab:cosmo_param}
\end{table}

The likelihood defined in Equation~\eqref{eq:likelihood} is then used to explore the parameter space $\Gamma$ via the public code \texttt{MontePython}~\cite{Brinckmann2018MontePython, Audren:2012wb} which uses a Metropolis-Hasting \ac{mcmc} algorithm for cosmological parameter estimation.
In this work, we stick to the standard $\Lambda$\ac{cdm} model, by sampling the following free parameters: the CDM physical density $\omega_\mathrm{cdm}$, the Hubble constant, $H_\mathrm{0}$, and the amplitude of the primordial power spectrum, $\ln(10^{10}A_\mathrm{s})$. 
The total matter abundance, $\Omega_\mathrm{m}$, and the linear amplitude of matter fluctuations, $\sigma_8$, are derived at each MCMC step.
In Table~\ref{tab:cosmo_param}, the fiducial (injected) values and prior choices for each parameter of the analysis are reported, together with the remaining three $\Lambda$CDM parameters - the baryon density, $\omega_{b}$, the primordial tilt $n_s$, and the redshift of reionization -- which we kept fixed to their fiducial values in all our runs.

To facilitate future multi-tracer analyses and guarantee the reproducibility of our findings, all underlying data and code are publicly released. 
The likelihood modules and the data generation pipeline developed for this work are openly available for the \texttt{MontePython} \ac{mcmc} samplers on GitHub\footnote{\url{https://github.com/MatteoSchulz/GWxHI_MultiCLASS}.}. 

\section{Results and Discussion}\label{sec:results}
Before discussing our results, one might want to visually verify the sensitivity of the cross-correlation signal to the underlying cosmological model.
\begin{figure}[ht]
    \centering
    \includegraphics[width=0.7\linewidth]{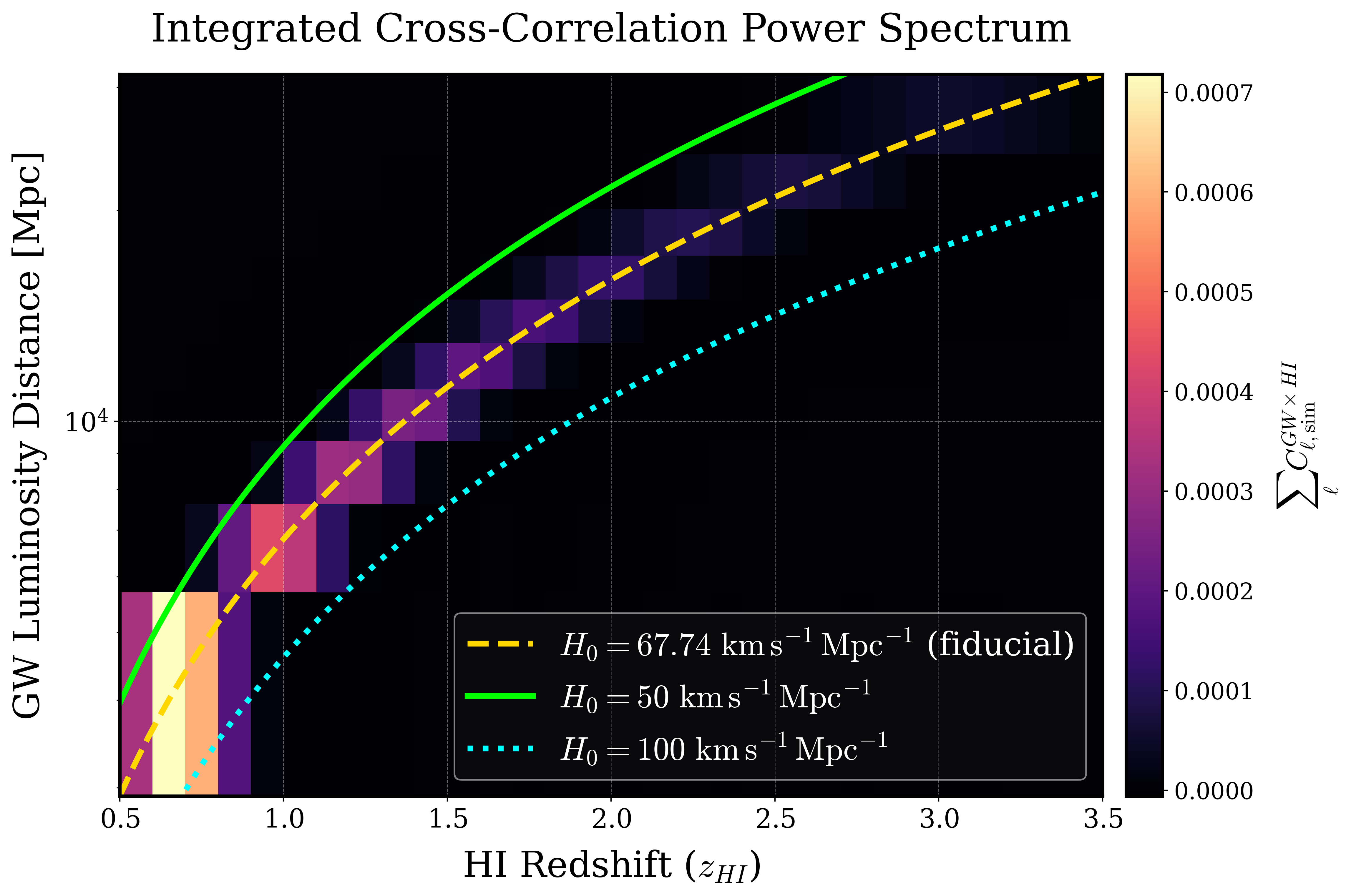} 
    \caption{Simulated cross-correlation power spectrum ($\sum_\ell \tilde{C}_\ell^{\rm HI, GW}$) integrated over the entire multipoles $\ell$ range, as a function of HI redshift and \ac{gw} luminosity distance.
    The overlaid lines represent the luminosity distance-redshift relation for the fiducial cosmology (dashed yellow) and for deviations: $H_\mathrm{0}=50$ (solid green) and $H_\mathrm{0}=100$ km s$^{-1}$ Mpc$^{-1}$ (dotted cyan).
    Note that the matter density, $\Omega_\mathrm{m}$, is fixed to its fiducial value in all three cases.
    The intensity of the diagonal band corresponds to the amplitude of the cross-correlation signal, which peaks where the exact physical overlap between the two tracers is achieved.}
    \label{fig:integrated_cls}
\end{figure}
This is done in Figure~\ref{fig:integrated_cls}, where we show the simulated cross-correlation power spectra ($\sum_\ell \tilde{C}_\ell^{\textrm{GW,H\textsc{i}}}$) integrated over the entire multipole range $\ell$, as a function of \ac{hi} redshift and \ac{gw} luminosity distance.
The intensity of the diagonal band corresponds to the level of cross-correlation between the two tracers.
As a reference, we overlay the theoretical luminosity distance-redshift relations, $d_L(z)$, for different values of the Hubble constant. 
The cross-correlation is maximized only when the correct distance-redshift relation is adopted, since the physical distances of the two tracers are assumed to be the same.
As expected, the fiducial (injected) model ($H_0 = 67.74$\,km\,s$^{-1}$\,Mpc$^{-1}$, dashed yellow line) aligns perfectly with the bright regions of maximal signal of the cross-correlation, while alternative $H_\mathrm{0}$ values (solid green line: $H_0 = 50 \text{ km s}^{-1} \text{Mpc}^{-1}$, and dotted cyan line: $H_0 = 100 \text{ km s}^{-1} \text{Mpc}^{-1}$) clearly fall into regions with weaker correlation, as the wrong $H_\mathrm{0}$ assumption breaks the spatial overlap between \ac{gw} and \ac{hi}.
This effect is the fundamental base on which all the cosmological implications are inferred in the following section.

\subsection{Comparison among different Gravitational Wave detector networks}
All our analyses are performed within the standard $\Lambda$\ac{cdm} framework. 
Since, within this model, the background expansion history is fully captured by the Hubble constant, $H_\mathrm{0}$, and the total fractional matter density, $\Omega_\mathrm{m}$, our baseline analysis varies only two parameters: $H_\mathrm{0}$ and the physical \ac{cdm} density, $\omega_\mathrm{cdm}$.
By keeping the baryon density $\omega_b$ fixed, varying $\omega_\mathrm{cdm}$ effectively allows us to map the full target parameter space of $H_\mathrm{0}$ and $\Omega_\mathrm{m}$.
All other cosmological parameters of $\Lambda$\ac{cdm} are kept fixed in the analysis.

To investigate the capabilities of the next-generation \ac{gw} detector in constraining cosmological parameters via cross-correlation, we start our analysis by evaluating the performance of three distinct networks introduced in Section~\ref{sec:gw_description}: ET-$\Delta$, ET-2L, and ET2L+CE.
Note that all the posteriors presented in this work are reported with their 1$\sigma$ and 2$\sigma$ contours.

\begin{figure}[ht]
    \centering
    \includegraphics[width=0.55\linewidth]{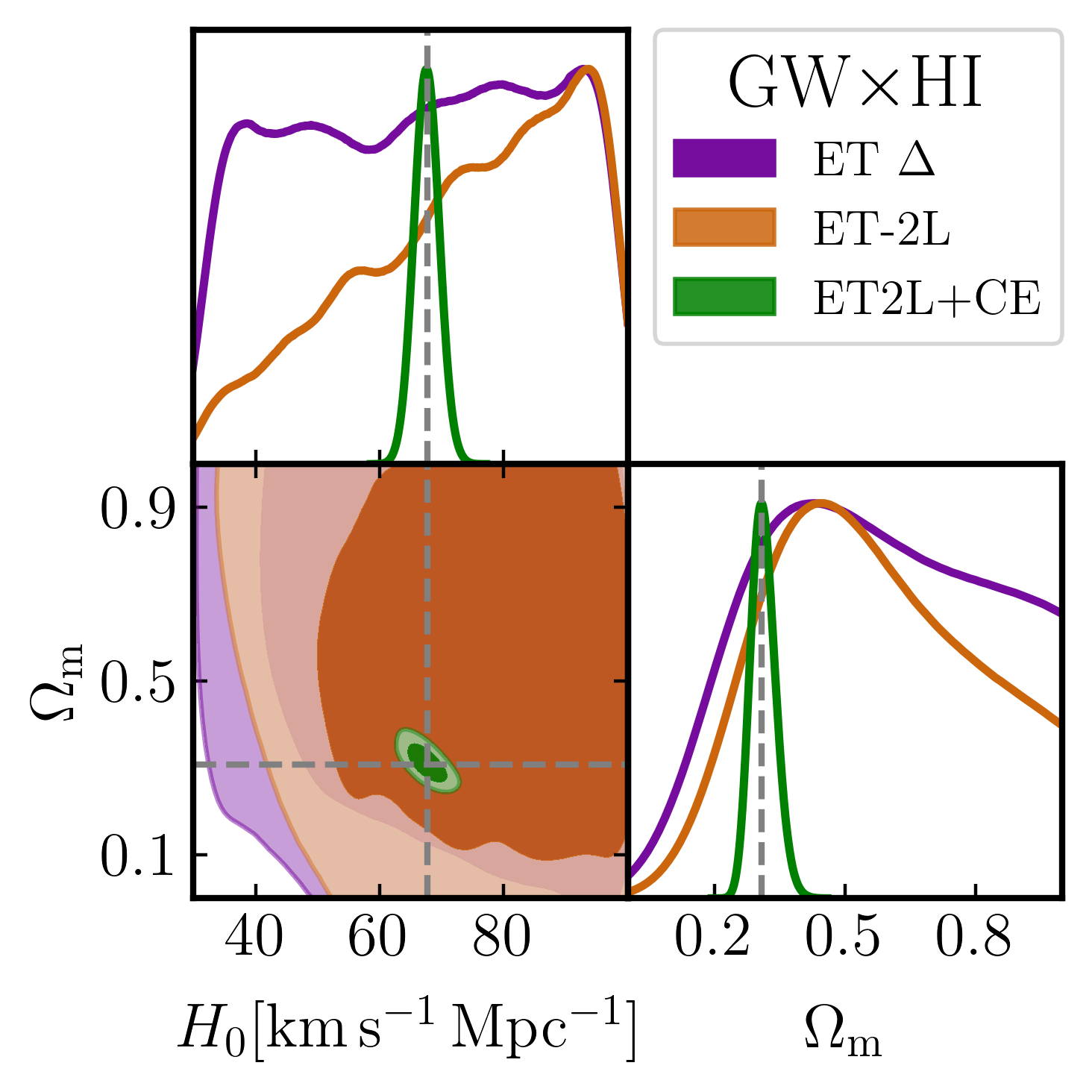}
    \caption{1D and 2D marginalized posteriors for $H_\mathrm{0}$ and $\Omega_\mathrm{m}$ evaluating the performance of different \ac{gw} detector networks when using only the \ac{gw}$\times$\ac{hi} cross-correlation signal.
    Specifically: ET\,$\Delta$ (purple) and ET-2L (orange), and ET2L+CE (green).
    Contours enclose the $68\%$ and $95\%$ confidence levels.
    Dashed lines mark the injected fiducial cosmology.}
    \label{fig:baseline_detectorComparison}
\end{figure}

In Figure~\ref{fig:baseline_detectorComparison} we show the 1D and 2D marginalized posteriors of these detectors when considering the \ac{gw}$\times$\ac{hi} cross-correlation term only.
Both the ET-$\Delta$ and the ET-2L configurations fail in constraining the background cosmology, as the posteriors are essentially uninformative for both $H_\mathrm{0}$ and $\Omega_\mathrm{m}$.
The main reason for this result is that both ET-$\Delta$ and ET-2L are single- or dual-site networks, thus lacking a large enough baseline to accurately triangulate the events and precisely resolve sky-localization and distance of the sources.
Nevertheless, since the ET-2L provides a larger baseline (and thus better localization capabilities) than the single-site ET-$\Delta$, it naturally provides marginally more informative posteriors.
This modest improvement justifies our choice of ET-2L as the baseline configuration to which we add CE for the main analysis.

The addition of CE to the ET-2L configuration forms a full three-detector network (ET2L+CE), which, thanks to its extremely large baseline, improves the observational capabilities, making it the only configuration for which we can extract stringent limits on the cosmological parameters even when only the \ac{gw}$\times$\ac{hi} contribution is included in the analysis.

It is important to stress that although the ``FullMatrix'' analysis would yield more stringent constraints, thanks to the inclusion the \ac{hi}$\times$\ac{hi} auto-correlation, it suffers from instrumental and astrophysical systematics linked to the foreground cleaning of radio observations.
In contrast, as the systematics of radio \ac{im} and \ac{gw} interferometry are uncorrelated, they cancel out in the \ac{gw}$\times$\ac{hi} cross-spectrum.
Therefore, the ability of the ET2L+CE network to constrain cosmological parameters using only this cross-correlation term represents an important proof of the potential of this technique.

\subsection{Constraints on the cosmic expansion history} \label{sec:baseline_analysis_AsFixed}
Having proven that the ET2L+CE setup gives the most constraining results in our analysis, we can focus on testing the robustness of the pipeline in recovering the background expansion parameters ($H_\mathrm{0}$, $\Omega_\mathrm{m}$).

\begin{figure}[ht]
    \centering
    \begin{subfigure}[b]{0.49\columnwidth}
        \centering
        \includegraphics[width=\linewidth]{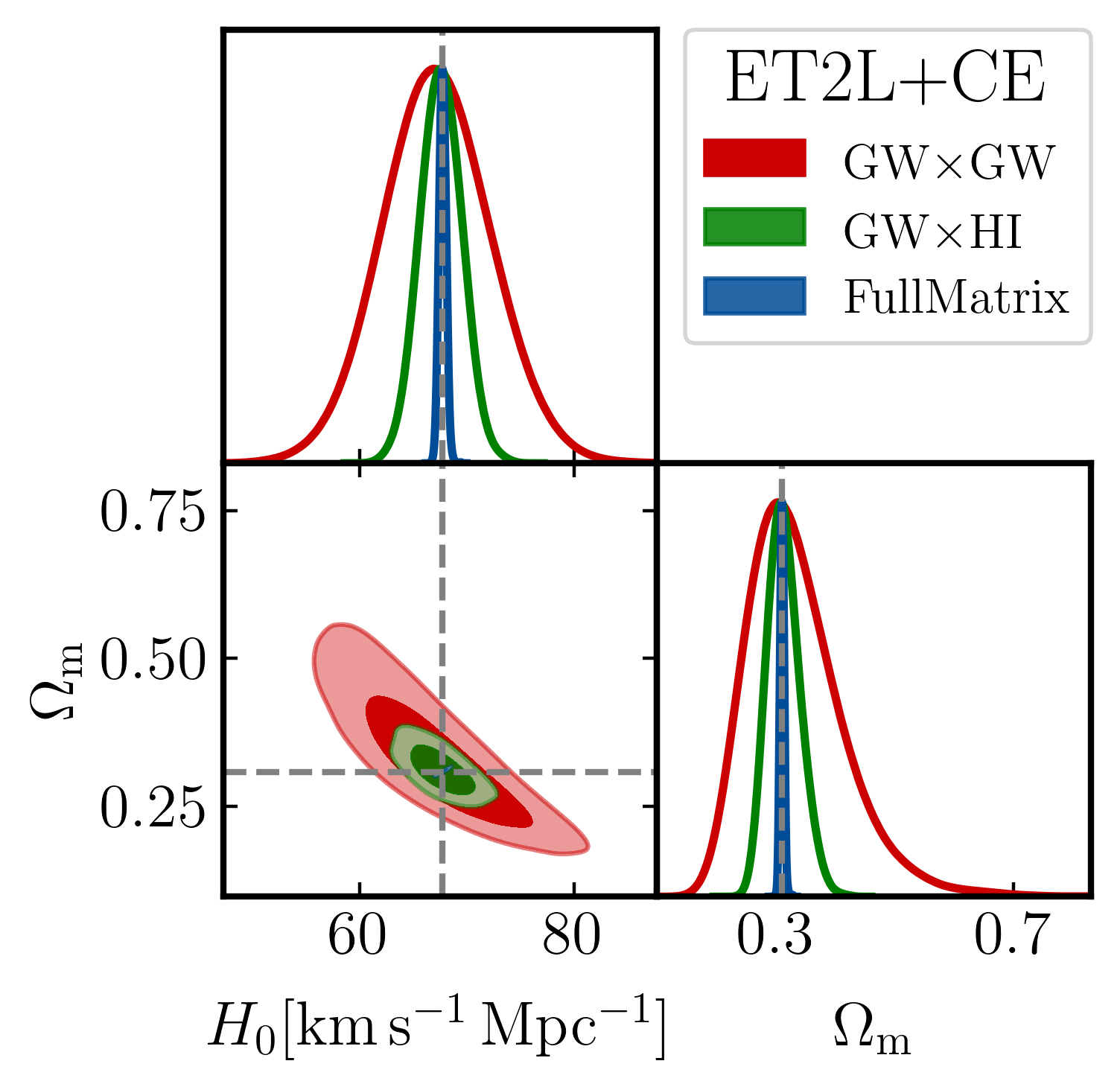} 
        \caption{}
        \label{fig:baseline_allParams}
    \end{subfigure}
    \hfill
    \begin{subfigure}[b]{0.49\columnwidth}
        \centering
        \includegraphics[width=\linewidth]{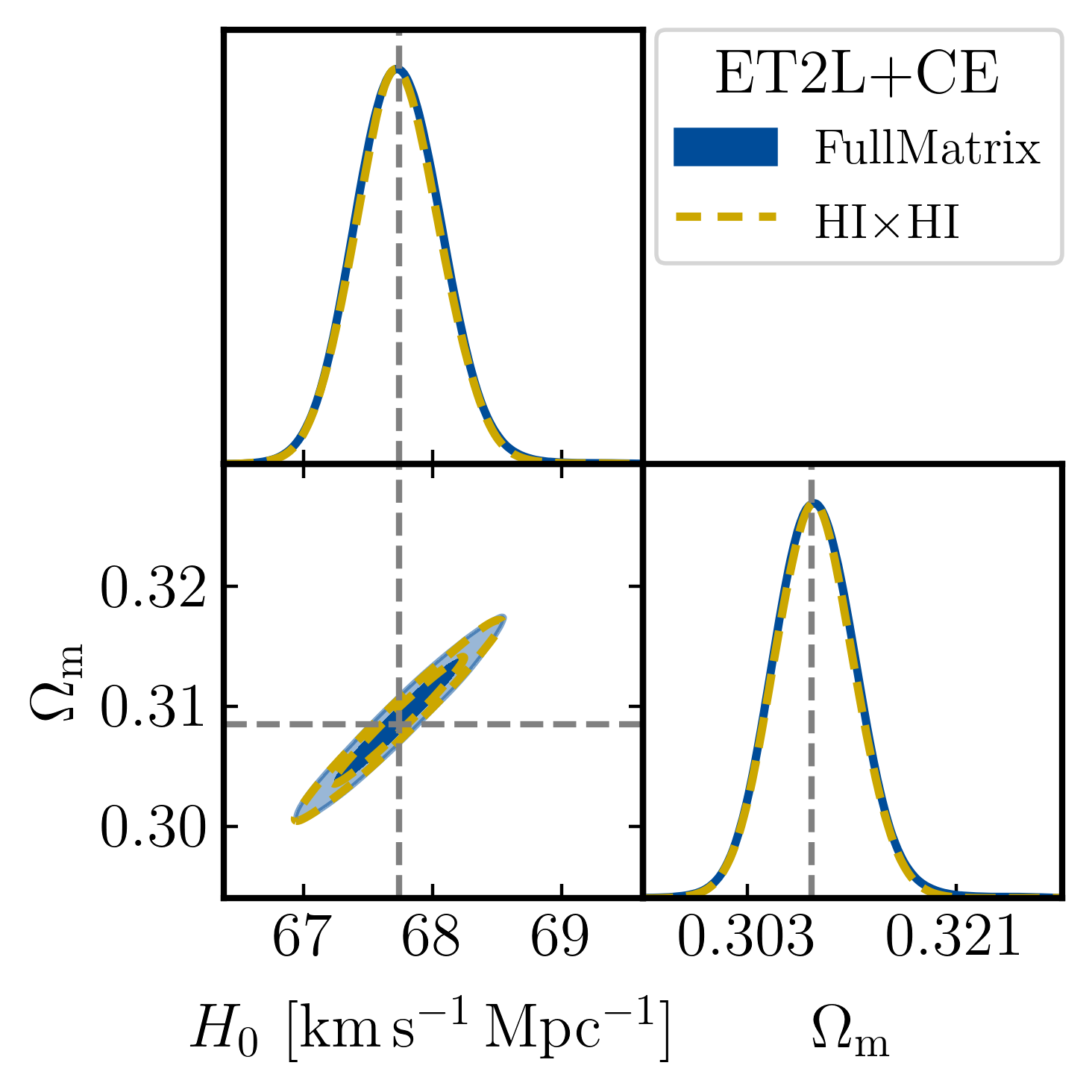} 
        \caption{}
        \label{fig:baseline_FullmatrHionly}
    \end{subfigure}
    \caption{1D and 2D marginalized posteriors for the baseline cosmological analysis ($H_\mathrm{0}$ and $\Omega_\mathrm{m}$) using the ET2L+CE network.
    Panel (a) constraints derived from the isolated \ac{gw}$\times$\ac{gw} auto-correlation (red), the \ac{gw}$\times$\ac{hi} cross-correlation (green), and the combined ``FullMatrix'' configuration (blue).
    Panel (b) compares the pure \ac{hi}$\times$\ac{hi} auto-correlation (dashed yellow) against the ``FullMatrix'' (blue). 
    Inner and outer contours enclose the $68\%$ and $95\%$ confidence levels, respectively. 
    Dashed lines indicate the injected fiducial parameter values.}
    \label{fig:baseline_mcmc}
\end{figure}

In Figure~\ref{fig:baseline_mcmc} we show the cosmological constraints from this baseline ($H_\mathrm{0}$, $\Omega_\mathrm{m}$) analysis.
It serves as a fundamental consistency check for our newly implemented methodology.
The pipeline successfully recovers the fiducial values of all injected cosmological parameters, confirming that both our modeling and our sampling frameworks do not introduce any additional bias in the analysis.

In Figure~\ref{fig:baseline_allParams} we show the impact of sequentially adding information on the overall constraining power.
In particular, the analysis including only the \ac{gw} auto-correlation (\ac{gw}$\times$\ac{gw}) is able to retrieve the injected cosmology, but it yields relatively broad constraints with respect to the runs which include \ac{hi} information, resulting in relative uncertainties of $\sim$\,$7.6\%$ for $H_\mathrm{0}$ and $\sim$\,$25\%$ for $\Omega_\mathrm{m}$.
Considering only the cross-correlation term (\ac{gw}$\times$\ac{hi}), the pipeline provides tighter constraints, narrowing the relative errors down to $\sim$\,$2.9\%$ and $\sim$\,$9\%$ for $H_\mathrm{0}$ and $\Omega_\mathrm{m}$, respectively.
This result is driven by the redshift information provided by \ac{hi}, which breaks the distance-redshift degeneracy intrinsic to \ac{gw} standard sirens.
By dropping the auto-correlation terms, the \ac{gw}$\times$\ac{hi} cross-correlation is free from any tracer-specific systematics.
Since noise and foregrounds are uncorrelated between the two distinct tracers, the isolated cross-correlation is an intrinsically clean and robust cosmological probe.
The ``FullMatrix'' case, which includes both the \ac{hi}$\times$\ac{hi} and \ac{gw}$\times$\ac{gw} auto-correlation spectra, as well as the \ac{gw}$\times$\ac{hi} cross-correlation term, provides the most stringent constraints on the considered cosmological parameters.
This combined analysis constraints $H_\mathrm{0}$ to a relative precision of $\sim$\,$0.5\%$ and $\Omega_\mathrm{m}$ to $\sim$\,$1.2\%$.

In Figure~\ref{fig:baseline_FullmatrHionly} we show a more detailed view on the constraints of the ``FullMatrix'' case compared to \ac{hi} auto-correlation (\ac{hi}$\times$\ac{hi}).
We can conclude that the great level of constraining power achieved in the ``FullMatrix'' analysis is fully driven by the \ac{hi} auto-correlation term, as the two posteriors completely overlap.
The high redshift precision of the \ac{hi} data over the entire redshift range removes any degeneracy linked to \ac{gw}s, thus making the analysis and the cosmological constraints independent of \ac{gw} data.
A more detailed discussion of the impact and role of \ac{hi} auto-correlation is presented in {Appendix}~\ref{sec:impactHI}.

\begin{figure}[ht]
    \centering
    \includegraphics[width=0.7\linewidth]{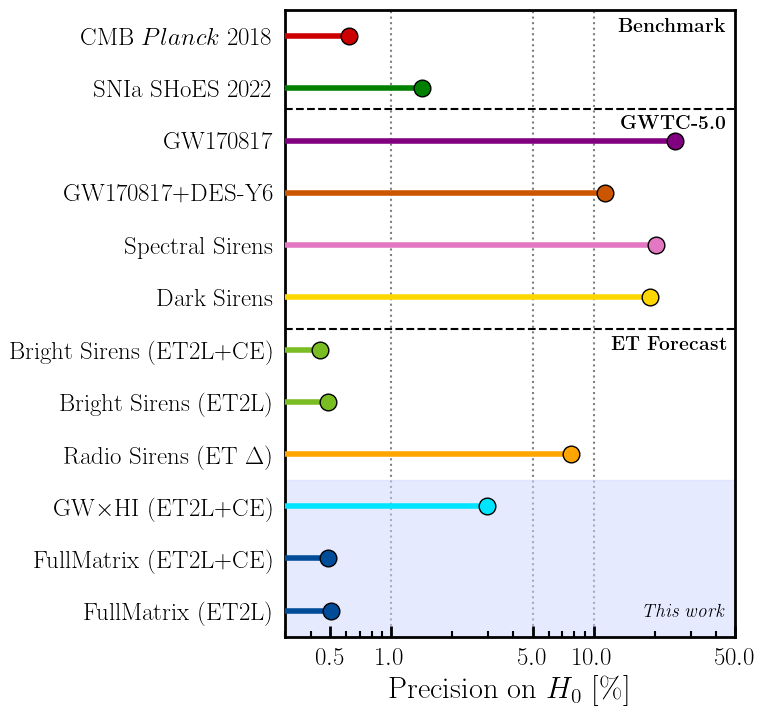}
    \caption{Compilation of $1\sigma$ fractional precision on $H_\mathrm{0}$ achieved by current measurements and future forecasts.
    The $x$-axis reports the percentage error on a logarithmic scale, with vertical dotted lines indicating 1\%, 5\%, and 10\% precision benchmarks.
    Horizontal dashed lines separate the data into three categories: benchmark constraints from \textit{Planck}\,2018 and SHoES (top), current-generation standard siren constraints from GWTC-5~\cite{lvk2026gwtc50_cosmo} (middle), and forecasts for next generation \ac{gw} observatories from Refs.~\cite{cozzumbo2024, dupletsa2026} alongside our results (bottom).
    Our measurements, derived from the \ac{gw}$\times$\ac{hi} cross-correlation (\textit{cyan}) and the ``FullMatrix'' approach (\textit{dark blue}) across different \ac{et} configurations, are highlighted at the bottom.}
    \label{fig:baseline_Planckcomparison}
\end{figure}

As we show in Figure~\ref{fig:baseline_Planckcomparison}, both the isolated cross-correlation and the ``FullMatrix'' configurations yield $H_\mathrm{0}$ constraints that are tighter than the corresponding results from current standard siren techniques, such as the bright siren GW170817 and the latest dark and spectral siren constraints from the recent GWTC-5.0 data release~\cite{lvk2026gwtc50_cosmo}.
In fact, our analysis allows for a sub-percent precision  for $H_\mathrm{0}$, that is competitive with the two fundamental cosmological probes: the early-Universe \textit{Planck}\,2018 CMB data~\cite{Aghanim2018PlanckVI} and the local SH0ES 2022 distance ladder~\cite{Riess2019Large}.
Our limits are competitive with other next-generation \ac{gw} detector forecasts based on standard siren techniques, such as \ac{bns} bright sirens~\cite{cozzumbo2024} and radio siren forecasts~\cite{dupletsa2026}.

Note also that, while both the \ac{gw}$\times$\ac{gw} and the \ac{gw}$\times$\ac{hi} cases clearly show a negative correlation, the ``FullMatrix'' case does break this degeneracy, resulting in a slight positive correlation between $H_\mathrm{0}$ and $\Omega_\mathrm{m}$.
This behavior is due to the inclusion of the \ac{hi}$\times$\ac{hi} term, which introduces different physical dependencies into the inference. 
On the one hand, the \ac{gw} measurement is primarily sensitive to the background expansion history \cite{schutz_1986, Holz:2005df}.
To preserve the distance-redshift relation, an increase in $H_\mathrm{0}$ must be compensated by a decrease in $\Omega_\mathrm{m}$,  which yields a negative $H_\mathrm{0}$-$\Omega_\mathrm{m}$ degeneracy.
On the other hand, the \ac{hi} auto-correlation is sensitive to the growth of structures and to the shape of the matter power spectrum, which have different scaling relations with $H_\mathrm{0}$ and $\Omega_\mathrm{m}$ \cite{Bull_2015}.
This additional sensitivity to the matter clustering scale breaks the background degeneracy, resulting in an overall rotation of the contours.

\subsection{Constraints on the amplitude of gravitational clustering}
Having established the robustness of the pipeline in recovering the background expansion parameters ($H_\mathrm{0}$, $\Omega_\mathrm{m}$), we expand the analysis to a larger set of cosmological parameters to test the sensitivity of this approach to physical observables associated with linear perturbations.
A major advantage of the \ac{gw}$\times$\ac{hi} cross-correlation is that, unlike conventional standard sirens methods, it naturally proves the large-scale distribution of matter perturbations.
Since the cross-correlation signal is driven by the clustering of both tracers within the cosmic web, it is supposed to be sensitive to the amplitude of the late-time linear matter power spectrum, as captured by the parameter $\sigma_8$, and in turn, to the primordial power spectrum amplitude, $A_\mathrm{s}$.

\begin{figure}[ht]
    \centering
    \begin{subfigure}[b]{0.49\columnwidth}
        \centering
        \includegraphics[width=\linewidth]{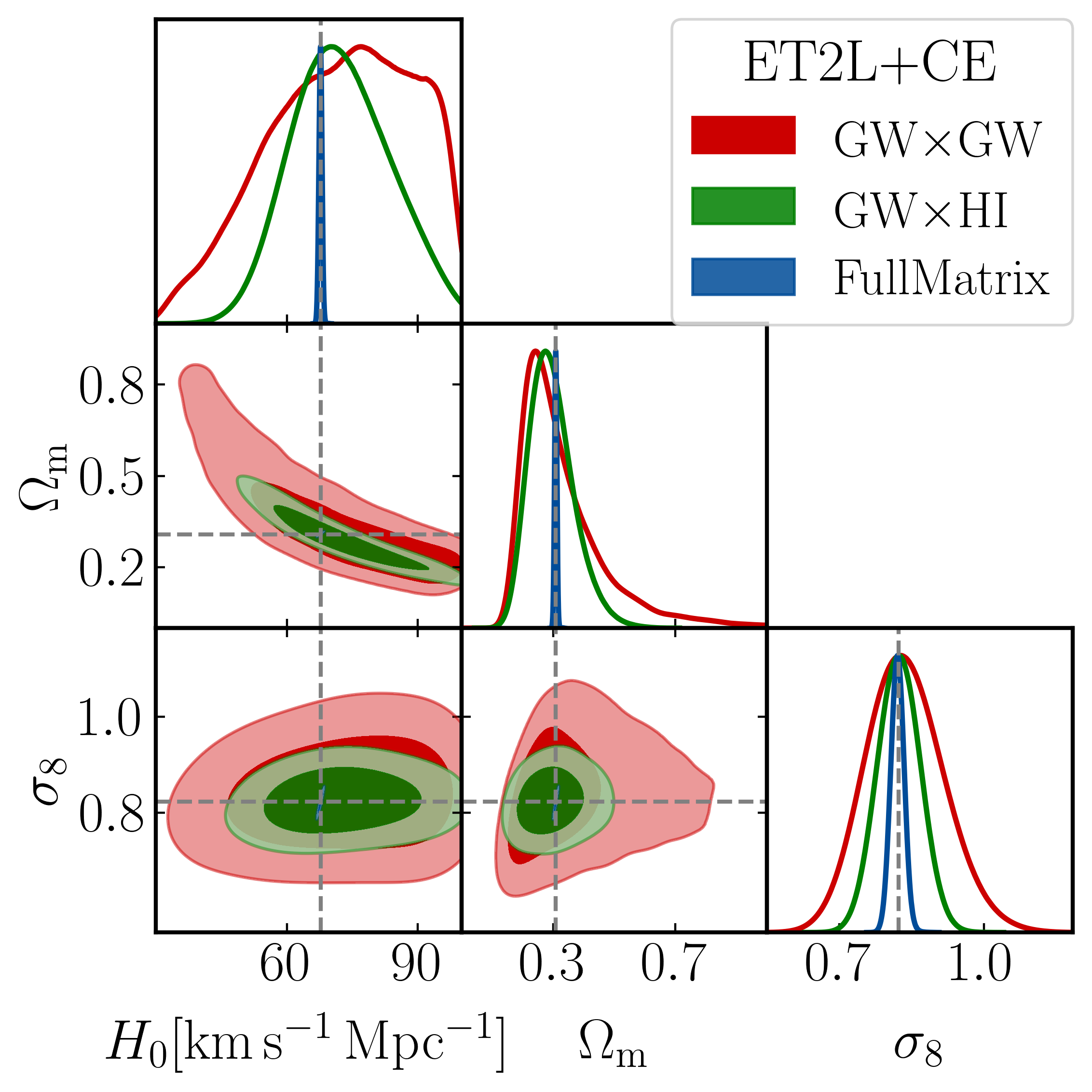} 
        \caption{}
        \label{fig:extended_subsetParam_comparison}
    \end{subfigure}
    \hfill
    \begin{subfigure}[b]{0.49\columnwidth}
        \centering
        \includegraphics[width=\linewidth]{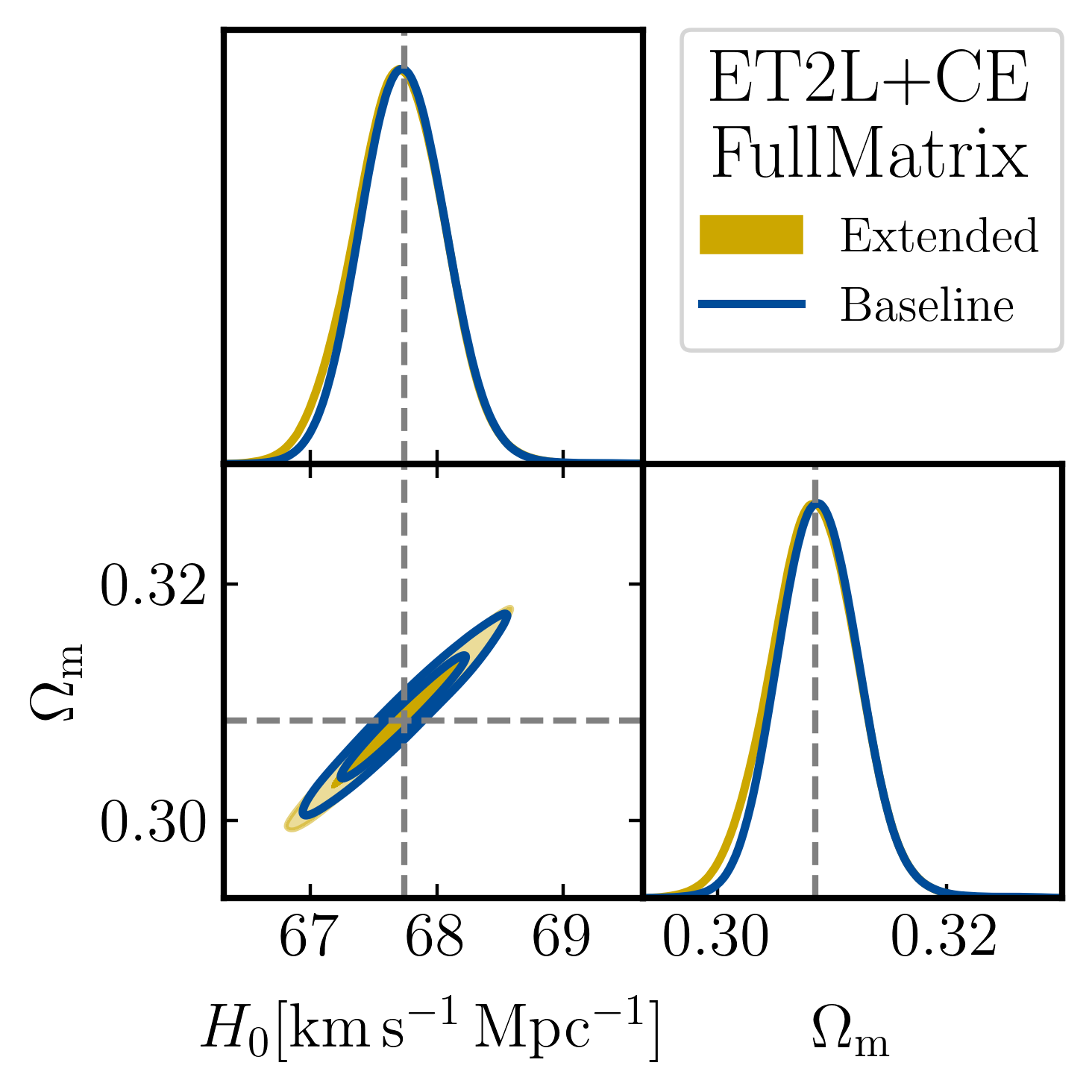} 
        \caption{}
        \label{fig:baseVSext_allParams}
    \end{subfigure}
    \caption{1D and 2D marginalized posteriors for the ET2L+CE network.
    Panel (a) displays the constraints on the extended parameter space ($H_\mathrm{0}$, $\Omega_\mathrm{m}$, and $\sigma_8$), comparing the isolated \ac{gw}$\times$\ac{gw} auto-correlation (red), the \ac{gw}$\times$\ac{hi} cross-correlation (green), and the combined ``FullMatrix'' (blue).
    Panel (b) compares the background constraints ($H_\mathrm{0}$ and $\Omega_\mathrm{m}$) derived from the ``FullMatrix'' configuration between the baseline analysis (fixed $A_\mathrm{s}$, blue) and the extended analysis (free $A_\mathrm{s}$, yellow).
    Contours enclose the $68\%$ and $95\%$ confidence levels, and dashed lines indicate the injected fiducial values.}
    \label{fig:extended_constraints}
\end{figure}

In Figure~\ref{fig:extended_subsetParam_comparison}, we plot the 1D and 2D marginalized posteriors for the analysis performed when adding $A_s$ as a free parameter, showing constraints on $H_\mathrm{0}$, $\Omega_\mathrm{m}$, and $\sigma_8$.
As in the results of our baseline case, the \ac{gw}$\times$\ac{gw} case alone provides insufficient constraining power, resulting in broad, uninformative posteriors with relative uncertainties of $\sim$\,$22.5\%$ for $H_\mathrm{0}$, $\sim$\,$48\%$ for $\Omega_\mathrm{m}$, and $\sim$\,$ 9.7\%$ for $\sigma_8$.
The isolated \ac{gw}$\times$\ac{hi} cross-correlation term provides a notable improvement, tightening the constraints considerably.
Relying only on cross-correlation, the relative errors shrink by roughly a factor of two, achieving $\sim$\,$15.8\%$ for $H_\mathrm{0}$, $\sim$\,$25.4\%$ for $\Omega_\mathrm{m}$, and $\sim$\,$5.3\%$ for $\sigma_8$.
However, note that freeing $A_\mathrm{s}$ degrades the constraints on $H_\mathrm{0}$ with respect to the fixed $A_\mathrm{s}$ case.
Specifically, the precision on $H_\mathrm{0}$ goes from $\sim$\,$2.9\%$ to $\sim$\,$15.8\%$, thus exposing the parameter degeneracy between the fluctuation amplitude and the background expansion.
As expected, when including the \ac{hi} auto-correlation and the \ac{gw}$\times$\ac{hi} cross-spectrum, the ``FullMatrix'' configuration successfully breaks this degeneracy and yields the tightest bounds on all investigated parameters. 
This combined approach achieves relative uncertainties of $\sim$\,$0.5\%$ and $\sim$\,$1.3\%$ for the background parameters $H_\mathrm{0}$ and $\Omega_\mathrm{m}$, alongside $\sim$\,$1.6\%$ precision on $\sigma_8$.

In Figure~\ref{fig:baseVSext_allParams} we compare the constraints on $H_\mathrm{0}$ and $\Omega_\mathrm{m}$ obtained in the baseline run of Section~\ref{sec:baseline_analysis_AsFixed} (fixed $A_\mathrm{s}$) to those from the extended run of this section (free $A_\mathrm{s}$).
Notably, such constraints do not show any sign of degradation, as the contours almost perfectly overlap between the two cases.
This result is a proof of the great potential of the \ac{et}-\ac{skao} synergy beyond background-only analyses generally performed in the context of standard sirens.
Our framework is robust and capable of simultaneously constraining the expansion history and the growth of structures at a very high degree of precision. Being able to simultaneously probe geometrical quantities and matter perturbations is crucial for testing the consistency across early- and late-universe probes within $\Lambda$CDM and beyond~\cite{Murgia:2020ryi,Abellan:2021sxk,DESI:2024mwx, descollaboration2026darkenergysurveyyear, Asgari_2021, Cozzumbo:2025ewt, Zhang_2005, Zhong_2023, Ruiz_Zapatero_2021}.

\section{Summary and Conclusion}\label{sec:conclusions}
We have presented a complete forecast for the cosmological constraining power of the cross-correlation between 21\,cm \ac{im} and \ac{gw}s for next-generation observatories, focusing on their optimal overlapping redshift range of $0.5 \le z \le 3.5$.
We have simulated the synergy between the \ac{skao} and various next-generation \ac{gw} detector networks involving \ac{et} and evaluated their joint potential in constraining both the expansion history and the large-scale structure of the Universe.
By using a tomographic angular power spectrum approach, we demonstrated that cross-correlating these two biased and independent tracers of the underlying dark matter distribution effectively breaks the distance-redshift degeneracy inherent to \ac{gw} standard sirens.
The cross-correlation signal is maximized when the exact physical spatial overlap between the two tracers is achieved, enabling a direct measurement of the luminosity distance-redshift relation.
Within the $\Lambda$CDM framework, this yields stringent constraints on $H_\mathrm{0}$ and $\Omega_\mathrm{m}$.
Furthermore, because this signal is driven by the clustering of the tracers within the cosmic web, it additionally allows us to map the amplitude of the underlying density fluctuations.
Ultimately, this methodology provides a powerful and independent probe of both the background expansion and the late-time growth of structures.

\begin{figure}[ht]
    \centering
    \begin{subfigure}[]{0.49\columnwidth}
        \centering
        \includegraphics[width=\linewidth]{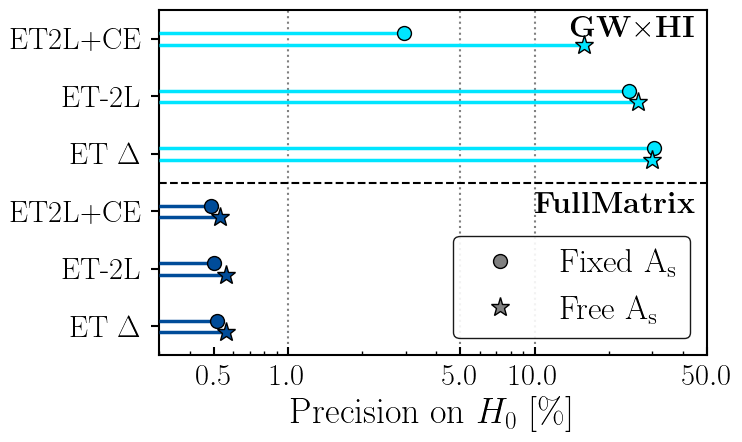} 
        \caption{}
        \label{fig:H0_precision_crossCorr_withExt}
    \end{subfigure}
    \hfill
    \begin{subfigure}[]{0.49\columnwidth}
        \centering
        \includegraphics[width=\linewidth]{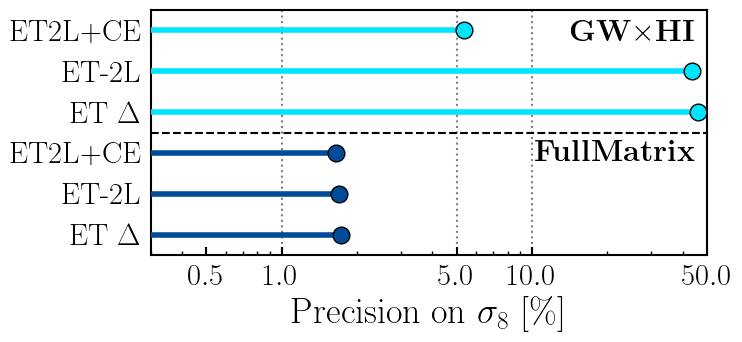} 
        \caption{}
        \label{fig:sigma_precision_crossCorr}
    \end{subfigure}
    \caption{Summary of the forecast $1\sigma$ precision on the cosmological parameters analyses in this work.
    The panels display the percentage error for \textit{(a)} the Hubble constant $H_\mathrm{0}$, and \textit{(b)} the clustering amplitude $\sigma_8$.
    The $x$-axis in each panel uses a logarithmic scale, with vertical dotted lines indicating 1\%, 5\%, and 10\% precision benchmarks.
    Results are categorized by methodology, separated by the horizontal dashed line: \ac{gw}$\times$\ac{hi} cross-correlation alone (\textit{cyan, top}) versus the ``FullMatrix'' analysis (\textit{dark blue, bottom}).
    For each network configuration (ET2L+CE, ET-2L, and ET\,$\Delta$), we compare our baseline analysis (with $A_s$ fixed, denoted by circles) against the extended one (with $A_s$ free, denoted by stars).}
    \label{fig:precision_crossCorr_withExt}
\end{figure}

Our main findings are reported in Figure~\ref{fig:precision_crossCorr_withExt}. They can be summarized as follows:
\begin{itemize}
    \item \textit{Constraints on the cosmic expansion history:} The ``FullMatrix'' analysis, which includes both the auto- and cross-correlation of \ac{gw}s and \ac{hi}, yields robust, percent-level constraints on background cosmology for the ET2L+CE network.
    We forecast a precision of approximately $0.5\%$ and $1.2\%$ for $H_\mathrm{0}$ and $\Omega_\mathrm{m}$, respectively.
    These results are competitive with current established cosmological probes, such as \textit{Planck} CMB~\cite{Aghanim2018PlanckVIII}.
    Additionally, this techniques has proven to match the precision of alternative \ac{gw} standard siren techniques applied to future \ac{gw} detectors~\cite{cozzumbo2024, dupletsa2026}.
    Remarkably, even when relying exclusively on the \ac{gw}$\times$\ac{hi} cross-correlation signal, the ET2L+CE configuration retains significant constraining power, yielding competitive measurements with a precision of $\sim$\,$2.9\%$ and $\sim$\,$9\%$ for $H_\mathrm{0}$ and $\Omega_\mathrm{m}$, respectively.
        
    \item \textit{Constraints on the amplitude of gravitational clustering:} By expanding the parameter space to treat the primordial power spectrum amplitude ($A_\mathrm{s}$) as a free parameter, our methodology effectively constrains late-time matter clustering.
    This is a key advantage with respect to conventional standard sirens techniques.
    In the ET2L+CE configuration, the ``FullMatrix'' analysis achieves an uncertainty of competitive precision of $\sim$\,$1.6\%$ on the late-time clustering amplitude $\sigma_8$.
    The corresponding analysis for \ac{gw}$\times$\ac{hi} yields a precision of $\sim$\,$5.3\%$ for $\sigma_8$.
    Introducing one more free parameter does not degrade the constraints on the background; the injected values of $H_\mathrm{0}$ and $\Omega_\mathrm{m}$ are still recovered at their stringent $\sim$\,$0.5\%$ and $\sim$\,$1.3\%$ precision levels.
\end{itemize}

We note that the constraints of the joint analysis are overwhelmingly driven by the \ac{hi} data.
Nevertheless, while the \ac{gw} data alone provide relatively weak cosmological bounds, its cross-correlation with the \ac{hi} ensures robustness against observational systematics while maintaining good overall constraining power.
While in this work we have focused on the specific synergy between \ac{et} and \ac{skao}, the methodologies and pipelines that we have developed are fully general and highly modular.
Our framework can be readily expanded to incorporate a wide variety of other cosmological tracers.
Future analyses could integrate cross-correlations with, e.g., photometric and spectroscopic galaxy surveys, high-redshift tracers like the Lyman-$\alpha$ forest, or emerging transient catalogs such as Fast Radio Bursts (FRBs).
Furthermore, as our pipeline is fully interfaced with publicly available cosmological software~\cite{Blas2011Cosmic, Brinckmann2018MontePython, Audren:2012wb, Torrado:2020dgo, mastrogiovanni2023icarogwpythonpackageinference, DupletsaHarms2023}, its applicability extends well beyond the baseline $\Lambda$\ac{cdm} model investigated here.
The framework is natively equipped to test a wide range of theoretical scenarios, enabling future studies to explore extended cosmologies, dynamical dark energy models, and non-parametric reconstructions of the cosmic expansion and structure growth history.
In fact, as an initial proof of concept for these beyond  $\Lambda$\ac{cdm} models, we tested the inclusion of the Chevallier–Polarski–Linder (CPL) parametrization for dynamical dark energy, and preliminary runs confirm that our precise constraining power on $H_\mathrm{0}$ and $\Omega_\mathrm{m}$ is fully preserved.
By providing this versatile multi-tracer theoretical and computational platform, we lay the groundwork for highly robust, cross-disciplinary cosmological analyses in the upcoming era of next-generation observatories.

\appendix

\section*{Acknowledgements}

The authors thank Julien Lesgourgues and Giulio Scelfo for insightful discussions.
The authors acknowledge support from the \href{https://www.sgws-community.eu}{Sardinian GW Science (SGWS) Community}: a joint program between the University of Cagliari and the Gran Sasso Science Institute (GSSI) promoted and coordinated by the cultural association \href{https://www.associazioneideas.eu}{IDeAS}.

\section*{Data availability}
To facilitate future multi-tracer analyses and guarantee the reproducibility of our findings, all underlying data and code are publicly released. 
The likelihood modules and the data generation pipeline developed for this work are openly available for the \texttt{MontePython} \ac{mcmc} samplers on GitHub: \url{https://github.com/MatteoSchulz/GWxHI_MultiCLASS}.

{\small \bibliography{main}}
\bibliographystyle{unsrt}

\newpage

\section{Relativistic Number Counts}\label{appendix:rel_n_counts}
In this Appendix we provide the complete expressions for the relativistic number counts effects introduced in equation \eqref{eq:angularFluctuations}:
\begin{equation}
    \begin{aligned}
    \Delta_{\ell,X}^\mathrm{den}(k,x) &= b_X \delta(k,\tau_x) j_\ell  \\
    \Delta_{\ell,X}^\mathrm{vel}(k,x) &=  \frac{k}{\mathcal{H}}j''_\ell  V(k,\tau_x) + \left[(f^\mathrm{evo}_X-3)\frac{\mathcal{H}}{k}j_\ell + \left(\frac{\mathcal{H}'}{\mathcal{H}^2}+\frac{2-5s_X}{\chi(x)\mathcal{H}}+5s_X-f^\mathrm{evo}_X\right)j'_\ell \right]  V(k,\tau_x)       \\
    \Delta_{\ell,X}^\mathrm{len}(k,x) &= \ell(\ell+1) \frac{2-5s_X}{2} \int_0^{\chi(x)} d\chi \frac{\chi(x)-\chi}{\chi(x) r} \left[\Phi(k,\tau_x)+\Psi(k,\tau_x)\right] j_\ell(k\chi)    \\
    \Delta_{\ell,X}^\mathrm{gr}(k,x)  &= \left[\left(\frac{\mathcal{H}'}{\mathcal{H}^2}+\frac{2-5s_X}{\chi(x)\mathcal{H}}+5s_X-f^\mathrm{evo}_X+1\right)\Psi(k,\tau_x) + \left(-2+5s_X\right) \Phi(k,\tau_x) + \mathcal{H}^{-1}\Phi'(k,\tau_x)\right] j_\ell + \\
    &+ \int_0^{\chi(x)} d\chi \frac{2-5s_X}{\chi(x)} \left[\Phi(k,\tau)+\Psi(k,\tau)\right]j_\ell(k\chi) \\
    &+ \int_0^{\chi(x)} d\chi \left(\frac{\mathcal{H}'}{\mathcal{H}^2}+\frac{2-5s_X}{\chi(x)\mathcal{H}}+5s_X-f^\mathrm{evo}_X\right)_{\chi(x)} \left[\Phi'(k,\tau)+\Psi'(k,\tau)\right] j_\ell(k\chi)
    \end{aligned}
\end{equation}
where all quantities introduced have following physical meaning: $b_X$ is the bias parameter, $s_X$ is the magnification bias parameter, $f^\mathrm{evo}_X$ is the evolution bias parameter, $\chi$ is the conformal distance on the light cone, $\tau=\tau_0-\chi$ is the conformal time, $\tau_x=\tau_0-\chi(x)$, $j_\ell$, $j'_\ell=\frac{dj_\ell}{dy}$, $j''_\ell=\frac{d^2j_\ell}{dy^2}$ are the Bessel functions and their derivatives (evaluated at $y=k\chi(x)$ when not explicitly stated), $\mathcal{H}$ is the conformal Hubble parameter, the prime symbol $'$ indicates derivatives with respect to conformal time, $\delta$ is the density contrast in the comoving gauge, $V$ is the peculiar velocity, $\Phi$ and $\Psi$ are Bardeen potentials (see e.g.,~reference \cite{Dio_2013} and references therein).

\section{Limber Approximation}
The most common approximation used to compute equation~\eqref{eq:cl_th} is the Limber approximation~\cite{Limber1953}, which, thanks to the oscillatory nature of the spherical Bessel functions, allows for the following simplification:
\begin{equation}
    j_\ell(k\chi) \approx \sqrt{\frac{\pi}{2\ell+1}} \delta_D \left[ \ell +\frac{1}{2} - k \chi \right]
\end{equation}
where $\delta_D$ is the Dirac delta function.
This allows for the use of the following asymptotic formula~\cite{LoVerde_2008}:
\begin{equation}
    \frac{2}{\pi}\int d k \,  k^2\, j_\ell(k\chi)\, j_\ell(k\chi')\simeq\frac{1}{\chi^2}\delta_D(\chi-\chi') +\mathcal{O}\left(\ell+1/2\right)^{-2}
\end{equation} 
and the evaluation of the power spectrum at the wavenumber projected on the angular scale $\ell$, i.e.:
\begin{equation}
    \mathcal{P}(k, z)=\mathcal{P}\left[k=\frac{\ell+1/2}{\chi(z)}, z\right] \, .
\end{equation}
By putting everything together, in the case of the density terms, we obtain:
\begin{equation}
    C_\ell^\mathrm{denXdenY} (x_i, x_j) = \int_{0}^{\infty}  \frac{c dz}{H(z) \chi^2(z)} \tilde{\mathcal{W}}_X(z, x_i) \tilde{\mathcal{W}}_Y(z, x_j) \mathcal{P} \left( k=\frac{\ell+1/2}{\chi(z)}, z \right)
    \label{eq:cl_th_limber}
\end{equation}
where $k$ is the comoving wavenumber, $\chi(z)$ is the comoving distance at redshift $z$, $H(z)$ is the Hubble parameter at $z$, $\tilde{\mathcal{W}}_X(z, x_i) =  \mathcal{J}_X(z)\, b_X(z)\, \mathcal{W}_{X}(z, x_i)\, H(z) /c$, with $\mathcal{J}_X(z)$ being the Jacobian of the distance, and $\mathcal{P}(k,z)$ is the primordial matter power spectrum.
Equation~\eqref{eq:cl_th_limber} is a much simpler integral to compute, as it does not involve the spherical Bessel functions nor double integration.
Note that this approximation is valid only if:
\begin{equation}
    \frac{\Delta k}{k} \gg \frac{2\pi}{\ell + 1/2}
\end{equation}
and
\begin{equation}
    \frac{\Delta \chi}{\chi} \gg \frac{2\pi}{\ell + 1/2}
\end{equation}
Physically, these conditions require that the functions varying along the line of sight (such as the window functions and the power spectrum) must be sufficiently broad and smooth to allow many rapid oscillations of the spherical Bessel functions.
In practice, this implies that the Limber approximation is highly accurate for broad redshift bins and small angular scales (high $\ell$), but it systematically fails for very narrow tomographic redshift bins or at large angular scales (low $\ell$), where the exact integration is strictly required.

\section{The impact of \textrm{H\textsc{i}} auto-correlation on the constraining power} \label{sec:impactHI}
As we discussed throughout the main text, it is crucial to precisely quantify the impact of the \ac{hi} contribution on the overall parameter estimation.
To isolate and evaluate the constraining power of \ac{hi}, we perform the analysis in a worsened observational setup, which we referred to as ``broad\ac{hi}''.
In this scenario, we double the width of the \ac{hi} redshift bins, thus halving the total number of tomographic bins, and we reduce the \ac{hi} redshift precision by a factor of two.
This intentional worsening of the configuration will result in a decrease in constraining power by \ac{hi}, which would help to distinguish between the different contributions to the overall constraint.
It is crucial to emphasize that the main results are obtained using the optimized baseline \ac{hi} configuration, and this degraded setup is introduced purely as a diagnostic tool.

\begin{figure}[ht]
    \centering
    \begin{subfigure}[b]{0.49\columnwidth}
        \centering
        \includegraphics[width=\linewidth]{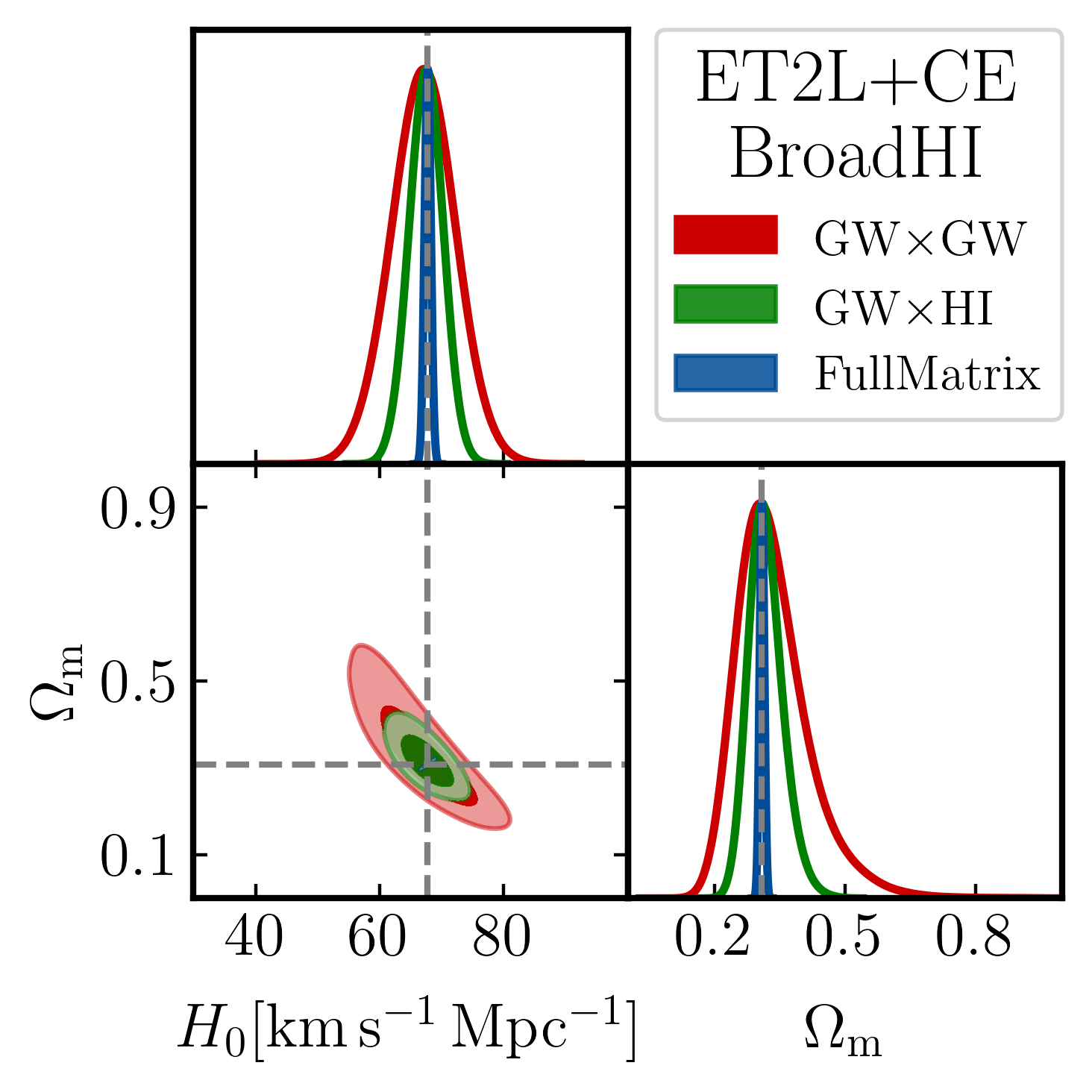} 
        \caption{}
        \label{fig:broadhi_allParams}
    \end{subfigure}
    \hfill
    \begin{subfigure}[b]{0.49\columnwidth}
        \centering
        \includegraphics[width=\linewidth]{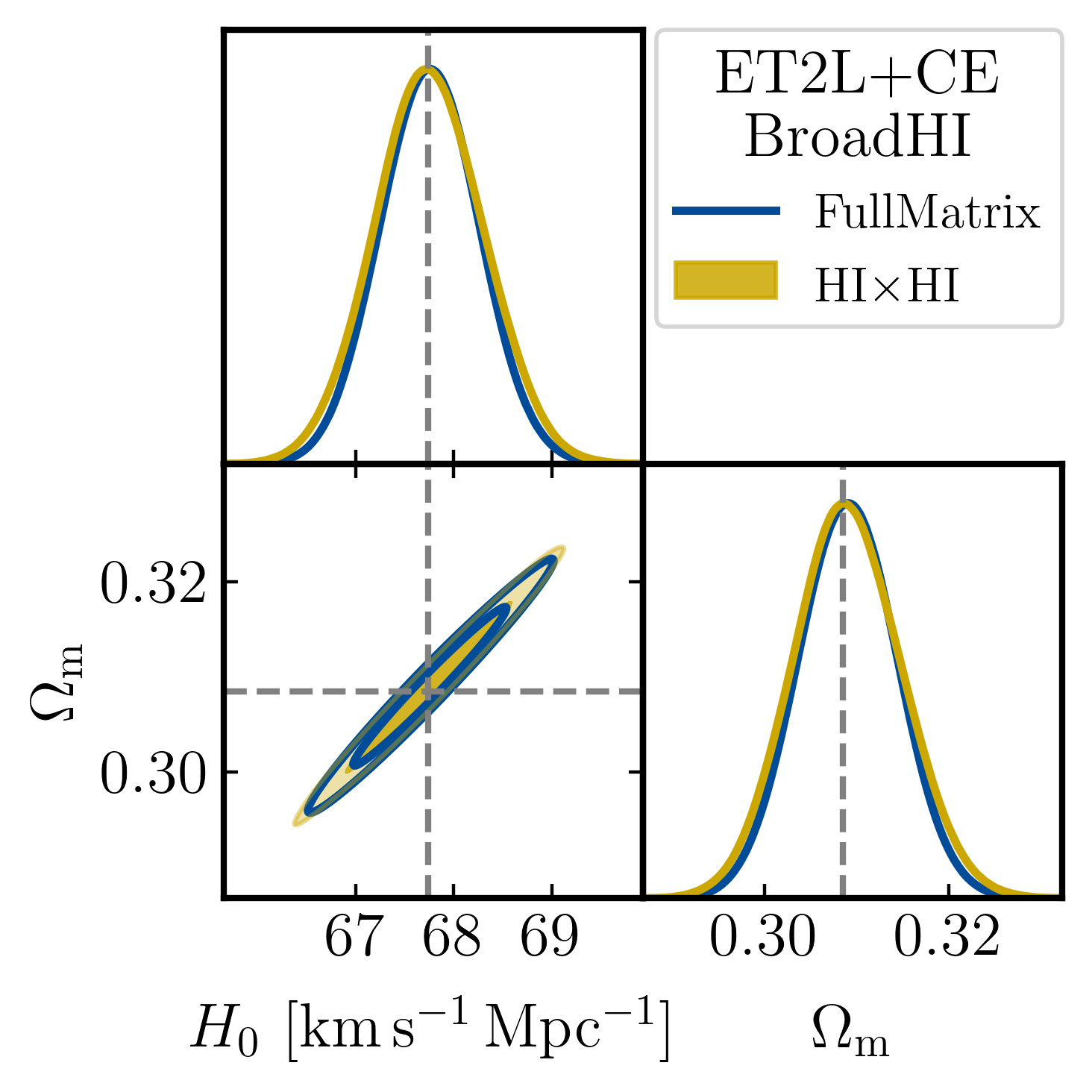} 
        \caption{}
        \label{fig:broadhi_FullmatrHionly}
    \end{subfigure}
    \caption{1D and 2D marginalized posteriors for $H_\mathrm{0}$ and $\Omega_\mathrm{m}$ evaluating the ET2L+CE network in the ``broad\ac{hi}'' configuration.
    Panel (a) compares the constraints derived from the isolated \ac{gw}$\times$\ac{gw} auto-correlation (red), the \ac{gw}$\times$\ac{hi} cross-correlation (green), and the combined ``FullMatrix'' (blue).
    Panel (b) compares the pure \ac{hi}$\times$\ac{hi} auto-correlation (yellow) against the ``FullMatrix'' (blue). Contours enclose the $68\%$ and $95\%$ confidence levels.
    Dashed lines mark the injected fiducial cosmology.}
    \label{fig:hi_fullmatr_comparison}
\end{figure}

In Figure~\ref{fig:broadhi_FullmatrHionly} we compare the constraints obtained from the \ac{hi} auto-correlation alone against the complete ``FullMatrix'' analysis in the ``broad\ac{hi}'' framework.
The pure \ac{hi}$\times$\ac{hi} constraints are only marginally larger than those derived from the ``FullMatrix'' case, yielding a precision of 0.8\% compared to 0.7\% of the ``FullMatrix'' analysis.
This confirms the assumption done in the main analysis: the high precision of \ac{hi} data drives the parameter estimation.
Nevertheless, the inclusion of \ac{gw} terms in the ``FullMatrix'' shows a slight tightening of the constraints, which suggests a small but non-negligible contribution from the second tracer.
This effect becomes relevant if the \ac{hi} observational capabilities assumed in this work are too optimistic or if the real data collection falls short of the ideal sensitivities simulated in this work.

\begin{figure}[ht]
    \centering
    \begin{subfigure}[b]{0.49\columnwidth}
        \centering
        \includegraphics[width=\linewidth]{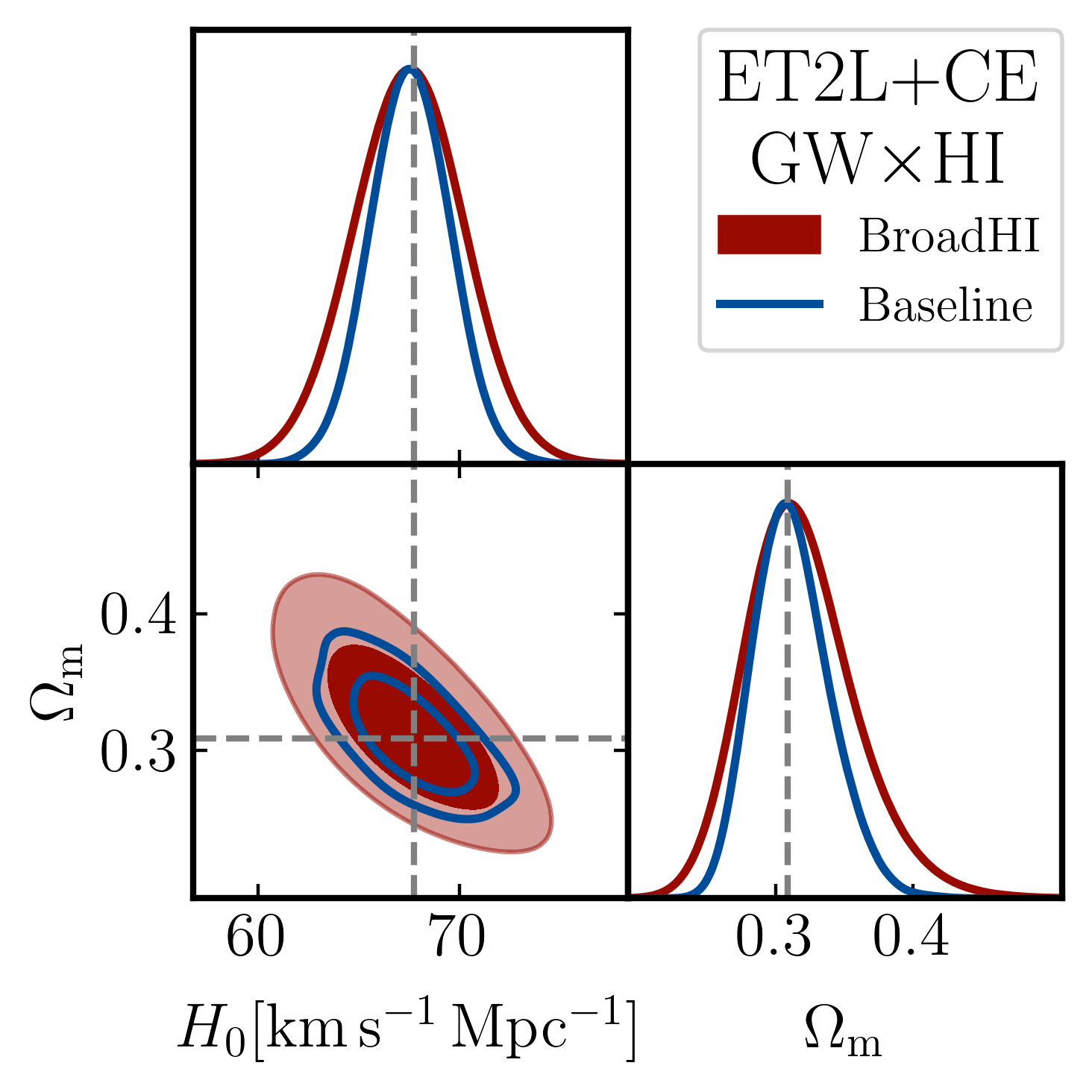}
        \caption{\ac{gw}$\times$\ac{hi}}
        \label{fig:broadhiHomcdm_comparison_GWHIcross}
    \end{subfigure}
    \hfill
    \begin{subfigure}[b]{0.49\columnwidth}
        \centering
        \includegraphics[width=\linewidth]{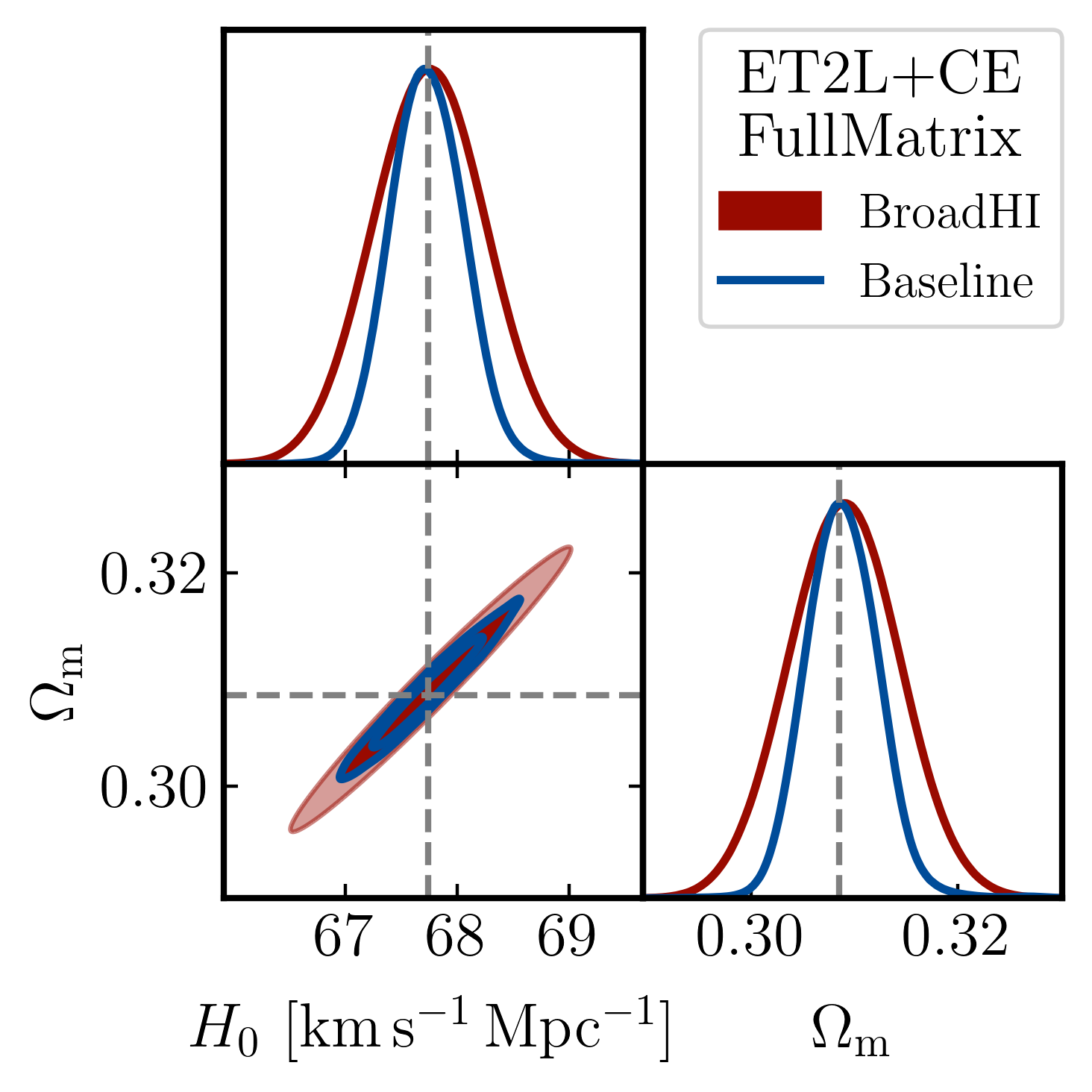} 
        \caption{``FullMatrix''}
        \label{fig:broadhiHomcdm_comparison_GWxHI}
    \end{subfigure}
    \caption{1D and 2D marginalized posteriors for $H_\mathrm{0}$ and $\Omega_\mathrm{m}$ evaluating the impact of reduced \ac{hi} tomographic resolution.
    Both panels compare the optimal baseline analysis (blue) against the degraded ``broad\ac{hi}'' framework (red) assuming the ET2L+CE network, focusing on the isolated \ac{gw}$\times$\ac{hi} cross-correlation (a), and the FullMatrix'' configuration (b). 
    Contours enclose the $68\%$ and $95\%$ confidence levels. Dashed lines mark the injected fiducial values.}
    \label{fig:broadhiHomcdm_comparison}
\end{figure}

Finally, it is crucial to quantify the worsening of the constraints due to the enlarging of the redshift bins and precision of the ``broad\ac{hi}'' analysis.
Figure~\ref{fig:broadhiHomcdm_comparison} shows this degradation for the 1D and 2D marginalized posteriors of $H_\mathrm{0}$ and $\Omega_\mathrm{m}$ in both the isolated cross-correlation signal (\ac{gw}$\times$\ac{hi}, panel a) and the ``FullMatrix'' case (panel b).
 As expected, the ``broad\ac{hi}'' configuration (red contours) yields systematically wider constraints compared to the baseline case (blue contours), with the relative errors in the cross-correlation alone degrading to from 3\% to 4\% for $H_\mathrm{0}$ and from 9\% to 12.5\% for $\Omega_\mathrm{m}$.
 In the ``FullMatrix'' case, constraints go from from 0.5\% to 0.7\% for $H_\mathrm{0}$ and from 1.2\% to 1.7\% for $\Omega_\mathrm{m}$.
As the power of cross-correlation comes from linking the \ac{gw} luminosity distance to the redshift scale of the second tracer, expanding and smearing the \ac{hi} tomographic redshift bins ``washes out'' the spatial overlap between the two tracers, thus reducing the overall cross-correlation signal, which ultimately results in weaker constraints.

These results demonstrate that the percent-level precision achieved by our optimal \ac{et}-\ac{skao} forecast heavily relies on the superb tomographic resolution capabilities of next-generation 21\,cm surveys.
Fine redshift binning and precision are critical not only to maximize pure \ac{hi} clustering information but also to optimize the spatial lock of cross-correlation with standard sirens.

\vfill

\end{document}

%% file: acronyms.tex
\acrodef{3g}[3G]{third-generation}

\acrodef{bh}[BH]{black hole}
\acrodef{bbh}[BBH]{binary black hole}
\acrodef{cbc}[CBC]{compact binary coalescence}
\acrodef{bns}[BNS]{binary neutron star}
\acrodef{cdm}[CDM]{cold dark matter}
\acrodef{cmb}[CMB]{cosmic microwave background}
\acrodef{de}[DE]{dark energy}
\acrodef{DEMNUni}[DEMNUni]{Dark Energy and Massive Neutrino Universe}
\acrodef{dm}[DM]{dark matter}
\acrodef{em}[EM]{electromagnetic}
\acrodef{eos}[EoS]{equation of state}
\acrodef{et}[ET]{Einstein Telescope}
\acrodef{fof}[FoF]{friend-of-friends}
\acrodef{gw}[GW]{gravitational wave}
\acrodef{hbi}[HBI]{hierarchical Bayesian inference}
\acrodef{hi}[\textrm{H\textsc{i}}]{neutral hydrogen}
\acrodef{im}[IM]{intensity mapping}
\acrodef{los}[LoS]{line-of-sight}
\acrodef{lss}[LSS]{large-scale structure}
\acrodef{pe}[PE]{parameter estimation}
\acrodef{skao}[SKAO]{SKA Observatory}
\acrodef{sne}[SNe]{supernovae}
\acrodef{snr}[SNR]{signal-to-noise ratio}
\acrodef{mcmc}[MCMC]{Markov chains Monte Carlo}

%% file: main.bbl
\begin{thebibliography}{100}

\bibitem{Riess1998Observational}
A.~G. Riess et~al.
\newblock {Observational evidence from supernovae for an accelerating universe and a cosmological constant}.
\newblock {\em Astron. J.}, 116:1009--1038, 1998.

\bibitem{Perlmutter1998Measurements}
S.~Perlmutter et~al.
\newblock {Measurements of $\Omega$ and $\Lambda$ from 42 high redshift supernovae}.
\newblock {\em Astrophys. J.}, 517:565--586, 1999.

\bibitem{Aghanim2018PlanckVI}
N.~Aghanim et~al.
\newblock {Planck 2018 results. VI. Cosmological parameters}.
\newblock {\em Astron. Astrophys.}, 641:A6, 2020.

\bibitem{Riess2019Large}
A.~G. Riess, S.~Casertano, W.~Yuan, L.~M. Macri, and D.~Scolnic.
\newblock {Large Magellanic Cloud Cepheid Standards Provide a 1\% Foundation for the Determination of the Hubble Constant and Stronger Evidence for Physics beyond $\Lambda$CDM}.
\newblock {\em The Astrophysical Journal}, 876(1):85, 2019.

\bibitem{Font_Ribera_2014}
A.~Font-Ribera et~al.
\newblock Quasar-lyman $\alpha$ forest cross-correlation from boss dr11: Baryon acoustic oscillations.
\newblock {\em Journal of Cosmology and Astroparticle Physics}, 2014(05):027–027, May 2014.

\bibitem{Verde2019Tensions}
L.~Verde, T.~Treu, and A.~G. Riess.
\newblock Tensions between the early and late universe.
\newblock {\em Nature Astronomy}, 3(10):891–895, September 2019.

\bibitem{DiValentino2021Realm}
Eleonora Di~Valentino, Olga Mena, Supriya Pan, Luca Visinelli, Weiqiang Yang, Alessandro Melchiorri, David~F. Mota, Adam~G. Riess, and Joseph Silk.
\newblock {In the realm of the Hubble tension\textemdash{}a review of solutions}.
\newblock {\em Class. Quant. Grav.}, 38(15):153001, 2021.

\bibitem{kamionkowski2022hubbletensionearlydark}
Marc Kamionkowski and Adam~G. Riess.
\newblock {The Hubble Tension and Early Dark Energy}, 2023.

\bibitem{hu2023hubbletensionevidencenew}
Jian-Ping Hu and Fa-Yin Wang.
\newblock Hubble tension: The evidence of new physics.
\newblock {\em Universe}, 9(2), 2023.

\bibitem{Poulin2018Early}
Vivian Poulin, Tristan~L. Smith, Tanvi Karwal, and Marc Kamionkowski.
\newblock {Early Dark Energy Can Resolve The Hubble Tension}.
\newblock {\em Phys. Rev. Lett.}, 122(22):221301, 2019.

\bibitem{Moresco:2022phi}
M.~Moresco et~al.
\newblock {Unveiling the Universe with emerging cosmological probes}.
\newblock {\em Living Rev. Rel.}, 25(1):6, 2022.

\bibitem{Abbott2017GW170817}
B.~Abbott, LIGO Scientific, Virgo Collaboration, KAGRA Collaboration, et~al.
\newblock {GW170817: Observation of Gravitational Waves from a Binary Neutron Star Inspiral}.
\newblock {\em Phys. Rev. Lett.}, 119:161101, Oct 2017.

\bibitem{Wong2019H0LiCOW}
K.~C. Wong et~al.
\newblock {H0LiCOW \textendash{} XIII. A 2.4 per cent measurement of H0 from lensed quasars: 5.3\ensuremath{\sigma} tension between early- and late-Universe probes}.
\newblock {\em Mon. Not. Roy. Astron. Soc.}, 498(1):1420--1439, 2020.

\bibitem{Freedman2019Carnegie}
Wendy~L. Freedman et~al.
\newblock {The Carnegie-Chicago Hubble Program. VIII. An Independent Determination of the Hubble Constant Based on the Tip of the Red Giant Branch}.
\newblock {\em The Astrophysical Journal}, 7 2019.

\bibitem{schutz_1986}
B.~F. {Schutz}.
\newblock {Determining the Hubble constant from gravitational wave observations}.
\newblock {\em Nature}, 323(6086):310--311, September 1986.

\bibitem{Ezquiaga:2022zkx}
Jose~Mar\'\i{}a Ezquiaga and Daniel~E. Holz.
\newblock {Spectral Sirens: Cosmology from the Full Mass Distribution of Compact Binaries}.
\newblock {\em Phys. Rev. Lett.}, 129(6):061102, 2022.

\bibitem{Chen:2024gdn}
H.-Y. Chen, J.~M. Ezquiaga, and I.~Gupta.
\newblock {Cosmography with next-generation gravitational wave detectors}.
\newblock {\em Class. Quant. Grav.}, 41(12):125004, 2024.

\bibitem{Markovic_1993cr}
Dragoljub Markovic.
\newblock {On the possibility of determining cosmological parameters from measurements of gravitational waves emitted by coalescing, compact binaries}.
\newblock {\em Phys. Rev. D}, 48:4738--4756, 1993.

\bibitem{dalal_2006}
Neal Dalal, Daniel~E. Holz, Scott~A. Hughes, and Bhuvnesh Jain.
\newblock Short grb and binary black hole standard sirens as a probe of dark energy.
\newblock {\em Phys. Rev. D}, 74:063006, Sep 2006.

\bibitem{LIGOScientific:2017adf}
B.~P. Abbott et~al.
\newblock {A gravitational-wave standard siren measurement of the Hubble constant}.
\newblock {\em Nature}, 551(7678):85--88, 2017.

\bibitem{Feeney:2018mkj}
Stephen~M. Feeney, Hiranya~V. Peiris, Andrew~R. Williamson, Samaya~M. Nissanke, Daniel~J. Mortlock, Justin Alsing, and Dan Scolnic.
\newblock {Prospects for resolving the Hubble constant tension with standard sirens}.
\newblock {\em Phys. Rev. Lett.}, 122(6):061105, 2019.

\bibitem{Palmese:2020}
A.~{Palmese}, et~al., and {DES Collaboration}.
\newblock {A Statistical Standard Siren Measurement of the Hubble Constant from the LIGO/Virgo Gravitational Wave Compact Object Merger GW190814 and Dark Energy Survey Galaxies}.
\newblock {\em \apjl}, 900(2):L33, September 2020.

\bibitem{Mancarella:2024qle}
Michele Mancarella, Francesco Iacovelli, Stefano Foffa, Niccol\`o Muttoni, and Michele Maggiore.
\newblock Accurate standard siren cosmology with joint gravitational-wave and $\ensuremath{\gamma}$-ray burst observations.
\newblock {\em Phys. Rev. Lett.}, 133:261001, Dec 2024.

\bibitem{cozzumbo2024}
Andrea {Cozzumbo}, Ulyana {Dupletsa}, Rodrigo {Calder{\'o}n}, Riccardo {Murgia}, Gor {Oganesyan}, and Marica {Branchesi}.
\newblock {Model-independent cosmology with joint observations of gravitational waves and {\ensuremath{\gamma}}-ray bursts}.
\newblock {\em \jcap}, 2025(5):021, May 2025.

\bibitem{Holz:2005df}
Daniel~E. Holz and Scott~A. Hughes.
\newblock {Using gravitational-wave standard sirens}.
\newblock {\em Astrophys. J.}, 629:15--22, 2005.

\bibitem{MacLeod:2007jd}
Chelsea~L. MacLeod and Craig~J. Hogan.
\newblock {Precision of Hubble constant derived using black hole binary absolute distances and statistical redshift information}.
\newblock {\em Phys. Rev. D}, 77:043512, 2008.

\bibitem{DelPozzo:2011vcw}
Walter Del~Pozzo.
\newblock {Inference of the cosmological parameters from gravitational waves: application to second generation interferometers}.
\newblock {\em Phys. Rev. D}, 86:043011, 2012.

\bibitem{Nishizawa:2016ood}
Atsushi Nishizawa.
\newblock {Measurement of Hubble constant with stellar-mass binary black holes}.
\newblock {\em Phys. Rev. D}, 96(10):101303, 2017.

\bibitem{Chen:2017rfc}
Hsin-Yu Chen, Maya Fishbach, and Daniel~E. Holz.
\newblock {A two per cent Hubble constant measurement from standard sirens within five years}.
\newblock {\em Nature}, 562(7728):545--547, 2018.

\bibitem{LIGOScientific:2018gmd}
M.~Fishbach et~al.
\newblock {A Standard Siren Measurement of the Hubble Constant from GW170817 without the Electromagnetic Counterpart}.
\newblock {\em Astrophys. J. Lett.}, 871(1):L13, 2019.

\bibitem{DES:2019ccw}
M.~Soares-Santos et~al.
\newblock {First Measurement of the Hubble Constant from a Dark Standard Siren using the Dark Energy Survey Galaxies and the LIGO/Virgo Binary\textendash{}Black-hole Merger GW170814}.
\newblock {\em Astrophys. J. Lett.}, 876(1):L7, 2019.

\bibitem{Gray:2019ksv}
Rachel Gray et~al.
\newblock {Cosmological inference using gravitational wave standard sirens: A mock data analysis}.
\newblock {\em Phys. Rev. D}, 101(12):122001, 2020.

\bibitem{LIGOScientific:2019zcs}
B.~P. Abbott et~al.
\newblock {A Gravitational-wave Measurement of the Hubble Constant Following the Second Observing Run of Advanced LIGO and Virgo}.
\newblock {\em Astrophys. J.}, 909(2):218, 2021.

\bibitem{Finke:2021aom}
Andreas Finke, Stefano Foffa, Francesco Iacovelli, Michele Maggiore, and Michele Mancarella.
\newblock {Cosmology with LIGO/Virgo dark sirens: Hubble parameter and modified gravitational wave propagation}.
\newblock {\em JCAP}, 08:026, 2021.

\bibitem{Abbott_2023}
R.~Abbott et~al.
\newblock Constraints on the cosmic expansion history from gwtc–3.
\newblock {\em The Astrophysical Journal}, 949(2):76, jun 2023.

\bibitem{Mancarella_2022_physRevD}
Michele Mancarella, Edwin Genoud-Prachex, and Michele Maggiore.
\newblock Cosmology and modified gravitational wave propagation from binary black hole population models.
\newblock {\em Phys. Rev. D}, 105:064030, Mar 2022.

\bibitem{Gair:2022zsa}
Jonathan~R. Gair et~al.
\newblock {The Hitchhiker\textquoteright{}s Guide to the Galaxy Catalog Approach for Dark Siren Gravitational-wave Cosmology}.
\newblock {\em Astron. J.}, 166(1):22, 2023.

\bibitem{Gray:2023wgj}
Rachel Gray et~al.
\newblock {Joint cosmological and gravitational-wave population inference using dark sirens and galaxy catalogues}.
\newblock {\em JCAP}, 12:023, 2023.

\bibitem{Borghi:2023opd}
Nicola Borghi, Michele Mancarella, Michele Moresco, Matteo Tagliazucchi, Francesco Iacovelli, Andrea Cimatti, and Michele Maggiore.
\newblock {Cosmology and Astrophysics with Standard Sirens and Galaxy Catalogs in View of Future Gravitational Wave Observations}.
\newblock {\em Astrophys. J.}, 964(2):191, 2024.

\bibitem{Bom:2024afj}
Clecio~R. Bom, V.~Alfradique, A.~Palmese, G.~Teixeira, L.~Santana-Silva, A.~Santos, and P.~Darc.
\newblock {A dark standard siren measurement of the Hubble constant following LIGO/Virgo/KAGRA O4a and previous runs}.
\newblock {\em Mon. Not. Roy. Astron. Soc.}, 535(1):961--975, 2024.

\bibitem{Borghi:2025nis}
Nicola Borghi et~al.
\newblock {Standard Sirens in 2040s: Probing the Cosmic Expansion History with Gravitational Waves and Spectroscopic Galaxy Surveys}.
\newblock {\em ArXiv e-prints}, dec 2025.
\newblock arXiv:2512.18369.

\bibitem{lvk2026gwtc50_cosmo}
The LIGO~Scientific Collaboration, the Virgo~Collaboration, and the KAGRA~Collaboration.
\newblock {GWTC-5.0}: Constraints on the cosmic expansion rate and modified gravitational-wave propagation, 2026.
\newblock arXiv:2605.27227.

\bibitem{Chernoff:1993th}
David~F. Chernoff and Lee~Samuel Finn.
\newblock {Gravitational radiation, inspiraling binaries, and cosmology}.
\newblock {\em Astrophys. J. Lett.}, 411:L5--L8, 1993.

\bibitem{Taylor:2012db}
Stephen~R. Taylor and Jonathan~R. Gair.
\newblock {Cosmology with the lights off: standard sirens in the Einstein Telescope era}.
\newblock {\em Phys. Rev. D}, 86:023502, 2012.

\bibitem{You:2020wju}
Zhi-Qiang You, Xing-Jiang Zhu, Gregory Ashton, Eric Thrane, and Zong-Hong Zhu.
\newblock {Standard-siren cosmology using gravitational waves from binary black holes}.
\newblock {\em Astrophys. J.}, 908(2):215, 2021.

\bibitem{Ye:2021klk}
Christine Ye and Maya Fishbach.
\newblock {Cosmology with standard sirens at cosmic noon}.
\newblock {\em Phys. Rev. D}, 104(4):043507, 2021.

\bibitem{Mastrogiovanni:2023emh}
Simone Mastrogiovanni, Danny Laghi, Rachel Gray, Giada~Caneva Santoro, Archisman Ghosh, Christos Karathanasis, Konstantin Leyde, Daniele~A. Steer, Stephane Perries, and Gregoire Pierra.
\newblock {Joint population and cosmological properties inference with gravitational waves standard sirens and galaxy surveys}.
\newblock {\em Phys. Rev. D}, 108(4):042002, 2023.

\bibitem{Ferraiuolo:2025evh}
S.~Ferraiuolo, S.~Mastrogiovanni, S.~Escoffier, and E.~Kajfasz.
\newblock {Inferring astrophysics and cosmology with individual compact binary coalescences and their gravitational-wave stochastic background}.
\newblock {\em Astron. Astrophys.}, 701:A36, 2025.

\bibitem{MaganaHernandez:2025cnu}
Ignacio Maga{\~n}a~Hernandez and Antonella Palmese.
\newblock {Spectral siren cosmology from gravitational-wave observations in GWTC-4.0}.
\newblock {\em ArXiv e-prints}, 9 2025.
\newblock arXiv:2509.03607.

\bibitem{Pierra:2026ffj}
Gr{\'e}goire Pierra and Alexander Papadopoulos.
\newblock {Heavy Black-Holes Also Matter in Standard Siren Cosmology}.
\newblock {\em ArXiv e-prints}, 2026.
\newblock arXiv:2601.03257.

\bibitem{Tagliazucchi:2026dpr}
Matteo Tagliazucchi, Michele Moresco, Alessandro Agapito, Michele Mancarella, Sarah Ferraiuolo, Simone Mastrogiovanni, Nicola Borghi, Francesco Pannarale, and Daniele Bonacorsi.
\newblock {Pushing spectral siren cosmology into the third-generation era: a blinded mock data challenge}.
\newblock {\em ArXiv e-prints}, 2 2026.

\bibitem{Bertheas:2026odj}
Tom Bertheas, Vasco Gennari, Dani{\`e}le Steer, and Nicola Tamanini.
\newblock {Spectral sirens cosmology from binary black holes populations with sharper mass features}.
\newblock {\em ArXiv e-prints}, 2026.
\newblock arXiv:2603.06792.

\bibitem{Diaz:2021pem}
C.~C. Diaz and S.~Mukherjee.
\newblock {Mapping the cosmic expansion history from LIGO-Virgo-KAGRA in synergy with DESI and SPHEREx}.
\newblock {\em Mon. Not. Roy. Astron. Soc.}, 511(2):2782--2795, 2022.

\bibitem{Scelfo:2021fqe}
G.~Scelfo, M.~Spinelli, A.~Raccanelli, L.~Boco, A.~Lapi, and M.~Viel.
\newblock {Gravitational waves \texttimes{} HI intensity mapping: cosmological and astrophysical applications}.
\newblock {\em JCAP}, 01(01):004, 2022.

\bibitem{Mukherjee:2022afz}
Suvodip Mukherjee, Alex Krolewski, Benjamin~D. Wandelt, and Joseph Silk.
\newblock Cross-correlating dark sirens and galaxies: Constraints on h0 from gwtc-3 of ligo–virgo–kagra.
\newblock {\em The Astrophysical Journal}, 975(2):189, nov 2024.

\bibitem{Mali:2024wpq}
Utkarsh Mali and Reed Essick.
\newblock {Striking a Chord with Spectral Sirens: Multiple Features in the Compact Binary Population Correlate with H$_{0}$}.
\newblock {\em Astrophys. J.}, 980(1):85, 2025.

\bibitem{Ferri:2024amc}
Jo{\~a}o Ferri, Ian~L. Tashiro, L.~Raul Abramo, Isabela Matos, Miguel Quartin, and Riccardo Sturani.
\newblock {A robust cosmic standard ruler from the cross-correlations of galaxies and dark sirens}.
\newblock {\em JCAP}, 04:008, 2025.

\bibitem{pedrotti2025}
A.~Pedrotti, M.~Mancarella, J.~Bel, and D.~Gerosa.
\newblock Cosmology with the angular cross-correlation of gravitational-wave and galaxy catalogs: forecasts for next-generation interferometers and the {Euclid} survey, 2025.
\newblock arXiv:2504.10482.

\bibitem{sala2025}
G.~Sala, A.~Cuoco, J.~Lesgourgues, K.-R. Revis, L.~{Valbusa Dall'Armi}, and S.~Casas.
\newblock Inferring cosmological parameters from galaxy and dark sirens cross-correlation.
\newblock {\em JCAP}, 2026(05):095, may 2026.

\bibitem{dematos2025}
Isabela {Santiago de Matos}, Charles {Dalang}, Tessa {Baker}, Raul {Abramo}, Jo{\~a}o {Ferri}, and Miguel {Quartin}.
\newblock {First measurement of the Hubble constant from gravitational wave-galaxy cross-correlations}, December 2025.
\newblock arXiv:2512.15380.

\bibitem{Dalang:2024gfk}
Charles Dalang, Bartolomeo Fiorini, and Tessa Baker.
\newblock {Large scale structure prior knowledge in the dark siren method}.
\newblock {\em JCAP}, 01:034, 2026.

\bibitem{Pan:2025iya}
Jiaming Pan, Dragan Huterer, Camille Avestruz, Damon H.~T. Cheung, Emery Trott, Neal Dalal, and Donghui Jeong.
\newblock {Determining the Hubble constant through cross-correlation of galaxies and gravitational waves}.
\newblock {\em Phys. Rev. D}, 113(10):103532, 2026.

\bibitem{Cross-Parkin:2026wyz}
Madeline~L. {Cross-Parkin}, Cullan {Howlett}, Leonardo {Giani}, Chris {Blake}, and Tamara~M. {Davis}.
\newblock {Dark siren cross-correlations and the sensitivity of $H_0$ to methodological choices}.
\newblock {\em arXiv e-prints}, page arXiv:2605.06783, May 2026.
\newblock arXiv:2605.06783.

\bibitem{Oguri_2016}
M.~Oguri.
\newblock Measuring the distance-redshift relation with the cross-correlation of gravitational wave standard sirens and galaxies.
\newblock {\em Physical Review D}, 93(8), April 2016.

\bibitem{Bera:2020}
Sayantani {Bera}, Divya {Rana}, Surhud {More}, and Sukanta {Bose}.
\newblock {Incompleteness Matters Not: Inference of H$_{0}$ from Binary Black Hole-Galaxy Cross-correlations}.
\newblock {\em \apj}, 902(1):79, October 2020.

\bibitem{mukherjee2018}
S.~Muherjee and B.D. Wandelt.
\newblock Beyond the classical distance-redshift test: cross-correlating redshift-free standard candles and sirens with redshift surveys, 2018.
\newblock arXiv:1808.06615.

\bibitem{kovetz2017lineintensitymapping2017status}
E.~D. Kovetz et~al.
\newblock Line-intensity mapping: 2017 status report, 2017.
\newblock arXiv:1709.09066.

\bibitem{Bernal_2022}
José~Luis Bernal and Ely~D. Kovetz.
\newblock Line-intensity mapping: theory review with a focus on star-formation lines.
\newblock {\em The Astronomy and Astrophysics Review}, 30(1), September 2022.

\bibitem{kovetz2019astrophysicscosmologylineintensitymapping}
Ely~D. Kovetz et~al.
\newblock {Astrophysics and Cosmology with Line-Intensity Mapping}.
\newblock {\em Bull. Am. Astron. Soc.}, 51(3):101, 2020.

\bibitem{santos2015cosmologyskahiintensity}
M.~G. Santos, P.~Bull, D.~Alonso, S.~Camera, P.~G. Ferreira, G.~Bernardi, R.~M., M.~Viel, F.~Villaescusa-Navarro, F.~B. Abdalla, M.~Jarvis, R.~B. Metcalf, A.~Pourtsidou, and L.~Wolz.
\newblock Cosmology with a {SKA HI} intensity mapping survey, 2015.
\newblock arXiv:1501.03989.

\bibitem{Bernal_2019}
José~Luis Bernal, Patrick~C. Breysse, and Ely~D. Kovetz.
\newblock Cosmic expansion history from line-intensity mapping.
\newblock {\em Physical Review Letters}, 123(25), December 2019.

\bibitem{chang_tc:2010}
T.-C. Chang, U.-L. Pen, K.~Bandura, and J.~B. Peterson.
\newblock {Conservative Constraints on Early Cosmology: an illustration of the Monte Python cosmological parameter inference code}.
\newblock {\em Nature}, 466:001, 2010.

\bibitem{Masui_2013}
K.~W. Masui, E.~R. Switzer, N.~Banavar, K.~Bandura, C.~Blake, L.-M. Calin, T.-C. Chang, X.~Chen, Y.-C. Li, Y.-W. Liao, A.~Natarajan, U.-L. Pen, J.~B. Peterson, J.~R. Shaw, and T.~C. Voytek.
\newblock Measurement of 21cm brightness fluctuations at z $\sim$ 0.8 in cross-correlation.
\newblock {\em The Astrophysical Journal Letters}, 763(1):L20, jan 2013.

\bibitem{anderson_2018}
C.~J. Anderson, N.~J. Luciw, Y.-C. Li, C.~Y. Kuo, J.~Yadav, K.~W. Masui, T.-C. Chang, X.~Chen, N.~Oppermann, Y.-W. Liao, U.-L. Pen, D.~C. Price, L.~Staveley-Smith, E.~R. Switzer, P.~T. Timbie, and L.~Wolz.
\newblock Low-amplitude clustering in low-redshift 21-cm intensity maps cross-correlated with 2df galaxy densities.
\newblock {\em Monthly Notices of the Royal Astronomical Society}, 476(3):3382--3392, 02 2018.

\bibitem{Wolz_2021}
L.~Wolz et~al.
\newblock H\textsc{i} constraints from the cross-correlation of eboss galaxies and green bank telescope intensity maps.
\newblock {\em Monthly Notices of the Royal Astronomical Society}, 510(3):3495–3511, December 2021.

\bibitem{chimecollaboration_2023}
CHIME Collaboration, Mandana Amiri, Kevin Bandura, Arnab Chakraborty, Matt Dobbs, Mateus Fandino, Simon Foreman, Hyoyin Gan, Mark Halpern, Alex~S. Hill, Gary Hinshaw, Carolin Höfer, T.~L. Landecker, Zack Li, Joshua MacEachern, Kiyoshi Masui, Juan Mena-Parra, Nikola Milutinovic, Arash Mirhosseini, Laura Newburgh, Anna Ordog, Sourabh Paul, Ue-Li Pen, Tristan Pinsonneault-Marotte, Alex Reda, J.~Richard Shaw, Seth~R. Siegel, Keith Vanderlinde, Haochen Wang, D.~V. Wiebe, and Dallas Wulf.
\newblock A detection of cosmological 21 cm emission from chime in cross-correlation with eboss measurements of the lyman-$\alpha$ forest, 2023.

\bibitem{Wang_2021}
J.~Wang, M.~G Santos, P.~Bull, K.~Grainge, S.~Cunnington, J.~Fonseca, M.~O. Irfan, Y.~Li, A.~Pourtsidou, P.~S. Soares, M.~Spinelli, G.~Bernardi, and B.~Engelbrecht.
\newblock H\textsc{i} intensity mapping with {MeerKAT}: calibration pipeline for multidish autocorrelation observations.
\newblock {\em Monthly Notices of the Royal Astronomical Society}, 505(3):3698–3721, May 2021.

\bibitem{meerKLASS_2025}
{MeerKLASS Collaboration}, Matilde {Barberi-Squarotti}, Jos{\'e}~L. {Bernal}, Philip {Bull}, Stefano {Camera}, Isabella~P. {Carucci}, Zhaoting {Chen}, Steven {Cunnington}, Brandon~N. {Engelbrecht}, Jos{\'e} {Fonseca}, Keith {Grainge}, Melis~O. {Irfan}, Yichao {Li}, Aishrila {Mazumder}, Sourabh {Paul}, Alkistis {Pourtsidou}, Mario~G. {Santos}, Marta {Spinelli}, Jingying {Wang}, Amadeus {Witzemann}, and Laura {Wolz}.
\newblock {MeerKLASS L-band deep-field intensity maps: entering the H I dominated regime}.
\newblock {\em \mnras}, 537(4):3632--3661, March 2025.

\bibitem{santos2017meerklassmeerkatlargearea}
M.~G. Santos et~al.
\newblock {MeerKLASS}: {MeerKAT} large area synoptic survey, 2017.
\newblock arXiv:1709.06099.

\bibitem{braun_2015}
R.~{Braun}, T.~{Bourke}, J.~A. {Green}, E.~{Keane}, and J.~{Wagg}.
\newblock {Advancing Astrophysics with the Square Kilometre Array}.
\newblock In {\em Advancing Astrophysics with the Square Kilometre Array (AASKA14)}, page 174, April 2015.

\bibitem{Bacon_Battye_2020}
D.~J. Bacon, R.~A. Battye, P.~Bull, S.~Camera, P.~G. Ferreira, I.~Harrison, D.~Parkinson, A.~Pourtsidou, M.~G. Santos, L.~Wolz, and et~al.
\newblock Cosmology with phase 1 of the square kilometre array red book 2018: Technical specifications and performance forecasts.
\newblock {\em Publications of the Astronomical Society of Australia}, 37:e007, 2020.

\bibitem{puntuto_2010}
M.~{Punturo} et~al.
\newblock {The Einstein Telescope: a third-generation gravitational wave observatory}.
\newblock {\em Classical and Quantum Gravity}, 27(19):194002, October 2010.

\bibitem{Punturo_2022}
M.~{Punturo} and {ET Collaboration Team}.
\newblock {Einstein Telescope and the 3rd generation GW observatories: science, technologies and perspectives}.
\newblock In {\em APS April Meeting Abstracts}, volume 2022 of {\em APS Meeting Abstracts}, page G07.002, April 2022.

\bibitem{Maggiore_2020}
M.~Maggiore, C.~Van~Den Broeck, N.~Bartolo, E.~Belgacem, D.~Bertacca, M.~A. Bizouard, M.~Branchesi, S.~Clesse, S.~Foffa, J.~García-Bellido, S.~Grimm, J.~Harms, T.~Hinderer, S.~Matarrese, C.~Palomba, M.~Peloso, A.~Ricciardone, and M.~Sakellariadou.
\newblock Science case for the {Einstein Telescope}.
\newblock {\em Journal of Cosmology and Astroparticle Physics}, 2020(03):050–050, March 2020.

\bibitem{Abac_2026}
A.~Abac and et~al. {(ET Collaboration)}.
\newblock {The Science of the Einstein Telescope}.
\newblock {\em Journal of Cosmology and Astroparticle Physics}, 2026(03):081, March 2026.

\bibitem{Branchesi_2023}
Marica Branchesi, Michele Maggiore, and et~al.
\newblock {Science with the Einstein Telescope: a comparison of different designs}.
\newblock {\em Journal of Cosmology and Astroparticle Physics}, 2023(07):068, July 2023.

\bibitem{dupletsa2026}
Ulyana Dupletsa, Simone Mastrogiovanni, Marta Spinelli, Tommaso Ronconi, Matteo Schulz, Riccardo Murgia, Jan Harms, Tessa Baker, Matteo Calabrese, Carmelita Carbone, Steven Cunnington, Ian Harrison, Konstantin Leyde, and Dounia Nanadoumgar-Lacroze.
\newblock Radio sirens: inferring {$H_0$} with binary black holes and neutral hydrogen in the era of the einstein telescope and the ska observatory, 2026.
\newblock arXiv:2605.12606.

\bibitem{switzer_2013}
E.~R. Switzer, K.~W. Masui, K.~Bandura, L.-M. Calin, T.-C. Chang, X.-L. Chen, Y.-C. Li, Y.-W. Liao, A.~Natarajan, U.-L. Pen, J.~B. Peterson, J.~R. Shaw, and T.~C. Voytek.
\newblock Determination of $z \sim 0.8$ neutral hydrogen fluctuations using the 21cm intensity mapping autocorrelation.
\newblock {\em Monthly Notices of the Royal Astronomical Society: Letters}, 434(1):L46--L50, 06 2013.

\bibitem{regos_1989}
E.~{Regos} and A.~S. {Szalay}.
\newblock {Multipole Expansion of the Large-Scale Velocity Field: Using the Tensor Window Function}.
\newblock {\em \apj}, 345:627, October 1989.

\bibitem{Scharf_1992}
C.~Scharf, Y.~Hoffman, O.~Lahav, and D.~Lynden-Bell.
\newblock Spherical harmonic analysis of iras galaxies: implications for the great attractor and cold dark matter.
\newblock {\em Monthly Notices of the Royal Astronomical Society}, 256(2):229--237, 05 1992.

\bibitem{Lahav_1994}
O.~{Lahav}, K.~B. {Fisher}, Y.~{Hoffman}, C.~A. {Scharf}, and S.~{Zaroubi}.
\newblock {Wiener Reconstruction of All-Sky Galaxy Surveys in Spherical Harmonics}.
\newblock {\em \apjl}, 423:L93, March 1994.

\bibitem{Fisher_1994}
K.~B. Fisher, C.~A. Scharf, and O.~Lahav.
\newblock A spherical harmonic approach to redshift distortion and a measurement of {$\Omega_0$} from the 1.2-{Jy IRAS Redshift Survey}.
\newblock {\em Monthly Notices of the Royal Astronomical Society}, 266(1):219--226, 01 1994.

\bibitem{menard2014clustering}
Brice Ménard, Ryan Scranton, Samuel Schmidt, Chris Morrison, Donghui Jeong, Tamas Budavari, and Mubdi Rahman.
\newblock Clustering-based redshift estimation: method and application to data, 2014.
\newblock arXiv: 1303.4722.

\bibitem{Challinor_2011}
Anthony Challinor and Antony Lewis.
\newblock Linear power spectrum of observed source number counts.
\newblock {\em Physical Review D}, 84(4), August 2011.

\bibitem{Bonvin_2011}
Camille Bonvin and Ruth Durrer.
\newblock What galaxy surveys really measure.
\newblock {\em Physical Review D}, 84(6), September 2011.

\bibitem{Limber1953}
D.~Nelson {Limber}.
\newblock {The Analysis of Counts of the Extragalactic Nebulae in Terms of a Fluctuating Density Field.}
\newblock {\em \apj}, 117:134, January 1953.

\bibitem{Dio_2013}
Enea~Di Dio, Francesco Montanari, Julien Lesgourgues, and Ruth Durrer.
\newblock The classgal code for relativistic cosmological large scale structure.
\newblock {\em Journal of Cosmology and Astroparticle Physics}, 2013(11):044–044, November 2013.

\bibitem{Bellomo20_multiclass}
Nicola Bellomo, Jos{\'{e}}~Luis Bernal, Giulio Scelfo, Alvise Raccanelli, and Licia Verde.
\newblock Beware of commonly used approximations. part i. errors in forecasts.
\newblock {\em Journal of Cosmology and Astroparticle Physics}, 2020(10):016--016, oct 2020.

\bibitem{schoeneberg_cls}
Nils {Sch{\"o}neberg}, Marko {Simonovi{\'c}}, Julien {Lesgourgues}, and Matias {Zaldarriaga}.
\newblock {Beyond the traditional line-of-sight approach of cosmological angular statistics}.
\newblock {\em \jcap}, 2018(10):047, October 2018.

\bibitem{DupletsaHarms2023}
U.~{Dupletsa}, J.~{Harms}, B.~{Banerjee}, M.~{Branchesi}, B.~{Goncharov}, A.~{Maselli}, A.~C.~S. {Oliveira}, S.~{Ronchini}, and J.~{Tissino}.
\newblock {GWFISH: A simulation software to evaluate parameter-estimation capabilities of gravitational-wave detector networks}.
\newblock {\em Astronomy and Computing}, 42:100671, January 2023.

\bibitem{Dupletsa:2024gfl}
Ulyana Dupletsa, Jan Harms, Ken K.~Y. Ng, Jacopo Tissino, Filippo Santoliquido, and Andrea Cozzumbo.
\newblock Validating prior-informed fisher-matrix analyses against gwtc data.
\newblock {\em Phys. Rev. D}, 111:024036, Jan 2025.

\bibitem{mastrogiovanni2023icarogwpythonpackageinference}
Simone Mastrogiovanni, Gr{\'e}goire Pierra, St{\'e}phane Perri{\`e}s, Danny Laghi, Giada Caneva~Santoro, Archisman Ghosh, Rachel Gray, Christos Karathanasis, and Konstantin Leyde.
\newblock {ICAROGW: A python package for inference of astrophysical population properties of noisy, heterogeneous, and incomplete observations}.
\newblock {\em Astronomy \& Astrophysics}, 682:A167, 2024.

\bibitem{Silverman1986}
Bernard~W. Silverman.
\newblock {\em Density Estimation for Statistics and Data Analysis}.
\newblock Chapman and Hall/CRC, London, 1986.

\bibitem{Scott1992}
David~W. Scott.
\newblock {\em Multivariate Density Estimation: Theory, Practice, and Visualization}.
\newblock John Wiley \& Sons, New York, 1992.

\bibitem{abbott2023population}
R.~Abbott et~al.
\newblock Population of merging compact binaries inferred using gravitational waves through gwtc-3.
\newblock {\em Physical Review X}, 13(1):011048, 2023.

\bibitem{Scelfo:2020jyw}
G.~Scelfo, L.~Boco, A.~Lapi, and M.~Viel.
\newblock {Exploring galaxies-gravitational waves cross-correlations as an astrophysical probe}.
\newblock {\em JCAP}, 10:045, 2020.

\bibitem{Libanore_2021}
S.~Libanore, M.~C. Artale, D.~Karagiannis, M.~Liguori, N.~Bartolo, Y.~Bouffanais, N.~Giacobbo, M.~Mapelli, and S.~Matarrese.
\newblock Gravitational wave mergers as tracers of large scale structures.
\newblock {\em Journal of Cosmology and Astroparticle Physics}, 2021(02):035–035, February 2021.

\bibitem{turner_1984}
E.~L. {Turner}, J.~P. {Ostriker}, and J.~R. {Gott}, III.
\newblock {The statistics of gravitational lenses : the distributions of image angular separations and lens redshifts.}
\newblock {\em \apj}, 284:1--22, September 1984.

\bibitem{Scelfo_2023}
Giulio Scelfo, Maria Berti, Alessandra Silvestri, and Matteo Viel.
\newblock Testing gravity with gravitational waves × electromagnetic probes cross-correlations.
\newblock {\em Journal of Cosmology and Astroparticle Physics}, 2023(02):010, February 2023.

\bibitem{greig_2017}
Bradley Greig and Andrei Mesinger.
\newblock The global history of reionization.
\newblock {\em Monthly Notices of the Royal Astronomical Society}, 465(4):4838--4852, 03 2017.

\bibitem{battye_2013}
R.~A. Battye, I.~W.~A. Browne, C.~Dickinson, G.~Heron, B.~Maffei, and A.~Pourtsidou.
\newblock {HI} intensity mapping: a single dish approach.
\newblock {\em Monthly Notices of the Royal Astronomical Society}, 434(2):1239--1256, 07 2013.

\bibitem{SKA:2018ckk}
{SKA Cosmology SWG}.
\newblock {Cosmology with Phase 1 of the Square Kilometre Array: Red Book 2018: Technical specifications and performance forecasts}.
\newblock {\em Publ. Astron. Soc. Austral.}, 37:e007, 2020.

\bibitem{Hall_2013}
A.~Hall, C.~Bonvin, and A.~Challinor.
\newblock Testing general relativity with 21-cm intensity mapping.
\newblock {\em Physical Review D}, 87(6), March 2013.

\bibitem{Alonso:2015sfa}
D.~Alonso and P.~G. Ferreira.
\newblock {Constraining ultralarge-scale cosmology with multiple tracers in optical and radio surveys}.
\newblock {\em Phys. Rev. D}, 92(6):063525, 2015.

\bibitem{crighton_2015}
N.~H.~M. Crighton, M.~T. Murphy, J.~X. Prochaska, G.~Worseck, M.~Rafelski, G.~D. Becker, S.~L. Ellison, M.~Fumagalli, S.~Lopez, A.~Meiksin, and J.~M. O'Meara.
\newblock The neutral hydrogen cosmological mass density at z = 5.
\newblock {\em Monthly Notices of the Royal Astronomical Society}, 452(1):217--234, 07 2015.

\bibitem{spinelli_2020}
M.~Spinelli, A~Zoldan, G.~De~Lucia, L.~Xie, and M.~Viel.
\newblock The atomic hydrogen content of the post-reionization universe.
\newblock {\em Monthly Notices of the Royal Astronomical Society}, 493(4):5434--5455, 03 2020.

\bibitem{Villaescusa-Navarro_2018}
F.~Villaescusa-Navarro, S.~Genel, E.~Castorina, An. Obuljen, D.~N. Spergel, L.~Hernquist, D.~Nelson, I.~P. Carucci, A.~Pillepich, F.~Marinacci, B.~Diemer, M.~Vogelsberger, R.~Weinberger, and R.~Pakmor.
\newblock Ingredients for 21 cm intensity mapping.
\newblock {\em The Astrophysical Journal}, 866(2):135, oct 2018.

\bibitem{Bull_2015}
P.~Bull, P.~G. Ferreira, P.~Patel, and M.~G. Santos.
\newblock Late-time cosmology with 21 cm intensity mapping experiments.
\newblock {\em The Astrophysical Journal}, 803(1):21, April 2015.

\bibitem{alonso_2014}
D.~Alonso, P.~Bull, P.~G. Ferreira, and M.~G. Santos.
\newblock Blind foreground subtraction for intensity mapping experiments.
\newblock {\em Monthly Notices of the Royal Astronomical Society}, 447(1):400--416, 12 2014.

\bibitem{Carucci_2020}
I.~P. Carucci, M.~O. Irfan, and J.~Bobin.
\newblock Recovery of 21-cm intensity maps with sparse component separation.
\newblock {\em Monthly Notices of the Royal Astronomical Society}, 499(1):304–319, September 2020.

\bibitem{Soares_2021}
P.~S. Soares, C.~A. Watkinson, S.~Cunnington, and A.~Pourtsidou.
\newblock Gaussian process regression for foreground removal in h\textsc{i} intensity mapping experiments.
\newblock {\em Monthly Notices of the Royal Astronomical Society}, 510(4):5872–5890, September 2021.

\bibitem{Matshawule_2021}
S.~D. Matshawule, M.~Spinelli, M.~G. Santos, and S.~Ngobese.
\newblock H\textsc{i} intensity mapping with meerkat: primary beam effects on foreground cleaning.
\newblock {\em Monthly Notices of the Royal Astronomical Society}, 506(4):5075–5092, June 2021.

\bibitem{Cunnington_2021}
S.~Cunnington, M.~O. Irfan, I.~P. Carucci, A.~Pourtsidou, and J.~Bobin.
\newblock 21-cm foregrounds and polarization leakage: cleaning and mitigation strategies.
\newblock {\em Monthly Notices of the Royal Astronomical Society}, 504(1):208–227, March 2021.

\bibitem{Ade2015PlanckXIII}
P.~A.~R. Ade et~al.
\newblock {Planck 2015 results. XIII. Cosmological parameters}.
\newblock {\em Astron. Astrophys.}, 594:A13, 2016.

\bibitem{Brinckmann2018MontePython}
Thejs Brinckmann and Julien Lesgourgues.
\newblock Montepython 3: Boosted mcmc sampler and other features.
\newblock {\em Physics of the Dark Universe}, 24:100260, 2019.

\bibitem{Audren:2012wb}
B.~Audren, J.~Lesgourgues, K.~Benabed, and S.~Prunet.
\newblock {Conservative Constraints on Early Cosmology: an illustration of the Monte Python cosmological parameter inference code}.
\newblock {\em JCAP}, 1302:001, 2013.

\bibitem{Murgia:2020ryi}
R.~Murgia, G.~F. Abell\'an, and V.~Poulin.
\newblock {Early dark energy resolution to the Hubble tension in light of weak lensing surveys and lensing anomalies}.
\newblock {\em Phys. Rev. D}, 103(6):063502, 2021.

\bibitem{Abellan:2021sxk}
Guillermo Franco~Abell{\'a}n, Riccardo Murgia, and Vivian Poulin.
\newblock {Linear cosmological constraints on two-body decaying dark matter scenarios and the S8 tension}.
\newblock {\em Phys. Rev. D}, 104(12):123533, 2021.

\bibitem{DESI:2024mwx}
A.~G. Adame et~al.
\newblock {DESI 2024 VI: Cosmological Constraints from the Measurements of Baryon Acoustic Oscillations}.
\newblock {\em Journal of Cosmology and Astroparticle Physics}, 2025(02):021, feb 2025.

\bibitem{descollaboration2026darkenergysurveyyear}
DES Collaboration, T.~M.~C. Abbott, et~al.
\newblock {Dark Energy Survey Year 6 Results: Cosmological Constraints from Galaxy Clustering and Weak Lensing}, 2026.
\newblock arXiv:2601.14559.

\bibitem{Asgari_2021}
Marika Asgari et~al.
\newblock Kids-1000 cosmology: Cosmic shear constraints and comparison between two point statistics.
\newblock {\em A\&A}, 645:A104, January 2021.

\bibitem{Cozzumbo:2025ewt}
Andrea Cozzumbo, Mattia Atzori~Corona, Riccardo Murgia, Maria Archidiacono, and Matteo Cadeddu.
\newblock {A short blanket for cosmology: the CMB lensing anomaly behind the preference for a negative neutrino mass}.
\newblock nov 2025.
\newblock arXiv:2511.01967.

\bibitem{Zhang_2005}
Jun Zhang, Lam Hui, and Albert Stebbins.
\newblock Isolating geometry in weak‐lensing measurements.
\newblock {\em The Astrophysical Journal}, 635(2):806–820, December 2005.

\bibitem{Zhong_2023}
Kunhao Zhong, Evan Saraivanov, Vivian Miranda, Jiachuan Xu, Tim Eifler, and Elisabeth Krause.
\newblock Growth and geometry split in light of the des-y3 survey.
\newblock {\em Physical Review D}, 107(12), June 2023.

\bibitem{Ruiz_Zapatero_2021}
Jamie Ruiz-Zapatero and et~al.
\newblock Geometry versus growth: Internal consistency of the flat {$\Lambda$CDM} model with kids-1000.
\newblock {\em Astronomy \& Astrophysics}, 655:A11, October 2021.

\bibitem{Aghanim2018PlanckVIII}
N.~Aghanim et~al.
\newblock {Planck 2018 results. VIII. Gravitational lensing}.
\newblock {\em Astron. Astrophys.}, 2018.

\bibitem{Blas2011Cosmic}
Diego Blas, Julien Lesgourgues, and Thomas Tram.
\newblock {The Cosmic Linear Anisotropy Solving System (CLASS) II: Approximation schemes}.
\newblock {\em JCAP}, 1107:034, 2011.

\bibitem{Torrado:2020dgo}
J.~Torrado and A.~Lewis.
\newblock {Cobaya: Code for Bayesian Analysis of hierarchical physical models}.
\newblock {\em JCAP}, 05:057, 2021.

\bibitem{LoVerde_2008}
Marilena LoVerde and Niayesh Afshordi.
\newblock Extended limber approximation.
\newblock {\em Physical Review D}, 78(12), December 2008.

\end{thebibliography}
